\documentclass[prx,twocolumn,superscriptaddress,amsmath,amssymb,amsthm,aps,nofootinbib]{revtex4-2} 
\usepackage{xcolor}
\usepackage{tikz}

\newcommand\irregularcircle[2]{
  \pgfextra {\pgfmathsetmacro\len{(#1)+rand*(#2)}}
  +(0:\len pt)
  \foreach \a in {10,20,...,350}{
    \pgfextra {\pgfmathsetmacro\len{(#1)+rand*(#2)}}
    -- +(\a:\len pt)
  } -- cycle
}

\usepackage{graphicx}
\usepackage{dcolumn}
\usepackage{bm}
\usepackage{dsfont}
\usepackage{hyperref}
\usepackage{comment}
\usepackage{float}

\newcommand{\prlsection}[1]{{\vspace{10 pt}\noindent \emph{{\textbf{#1}}.}}}

\newcommand{\spec}{\operatorname{spec}}

\usepackage{braket}

\newtheorem{proposition}{Proposition}

\newtheorem{remark}{Remark}

\newcommand{\id}{\mathds{1}}

\begin{document}

\title{Overlapping qubits from non-isometric maps and de Sitter tensor networks}

\author{ChunJun Cao}
\affiliation{Joint Center for Quantum Information and Computer Science, University of Maryland, College Park, MD, 20742, USA.}
 \affiliation{Institute for Quantum Information and Matter
  California Institute of Technology,
  1200 E California Blvd, Pasadena, CA 91125, USA.}
 \affiliation{Department of Physics, Virginia Tech, Blacksburg, VA, 24061, USA}

\author{Wissam Chemissany}
\affiliation{%
David Rittenhouse Laboratory, University of Pennsylvania, Philadelphia, PA 19104, USA 
}%

\author{Alexander Jahn}
\affiliation{%
 Institute for Quantum Information and Matter
  California Institute of Technology,
  1200 E California Blvd, Pasadena, CA 91125, USA.}%
\affiliation{%
 Department of Physics
 Freie Universit\"at Berlin
 14195 Berlin, Germany.}

\author{Zolt\'an Zimbor\'as}
\affiliation{
 Wigner Research Centre for Physics of the Hungarian Academy of Sciences, Budapest, Hungary
}%

\begin{abstract}

We construct approximately local observables, or ``overlapping qubits'', using non-isometric maps and show that processes in local effective theories can be spoofed with a quantum system with fewer degrees of freedom, similar to our expectation in holography. Furthermore, the spoofed system naturally deviates from an actual local theory in ways that can be identified with features in quantum gravity. For a concrete example, we construct two MERA toy models of de Sitter space-time and explain how the exponential expansion in global de Sitter can be spoofed with many fewer quantum degrees of freedom and that local physics may be approximately preserved for an exceedingly long time before breaking down. We highlight how approximate overlapping qubits are conceptually connected to Hilbert space dimension verification, degree-of-freedom counting in black holes and holography, and approximate locality in quantum gravity.
\end{abstract}

\maketitle

\section{Introduction} 
Although we often take for granted that the Hilbert space of $N$ qubits should be $2^N$-dimensional, the qubits we encounter in experiments are mere approximations of the ideal. In practice, the actual Hilbert space dimension of such physical qubits can be difficult to verify with limited computational resources and finite experimental precision. However, this verification problem is not to be lightly dismissed as a nuisance as it carries profound physical consequences even in theory. For example, in Ref.~\cite{chao2017overlapping} it was shown that if we only require certain quantum measurement outcomes to be reproduced approximately, it is possible to spoof even $N=O(\exp(\epsilon^2 n))$ qubits using only $n$ exact qubits as long as the $N$ qubits can overlap; that is, elements of the Pauli algebra of any two such qubits $P_i, P_j$ satisfy $\Vert [P_i,P_j]\Vert <\epsilon$. 
This spoofing is not without its limitations; for example, Ref.\ \cite{chao2017overlapping} devised a verification protocol with $\mathrm{poly}(N)$ complexity.

\begin{figure}[ht]
    \centering
    \includegraphics[width=0.35\textwidth]{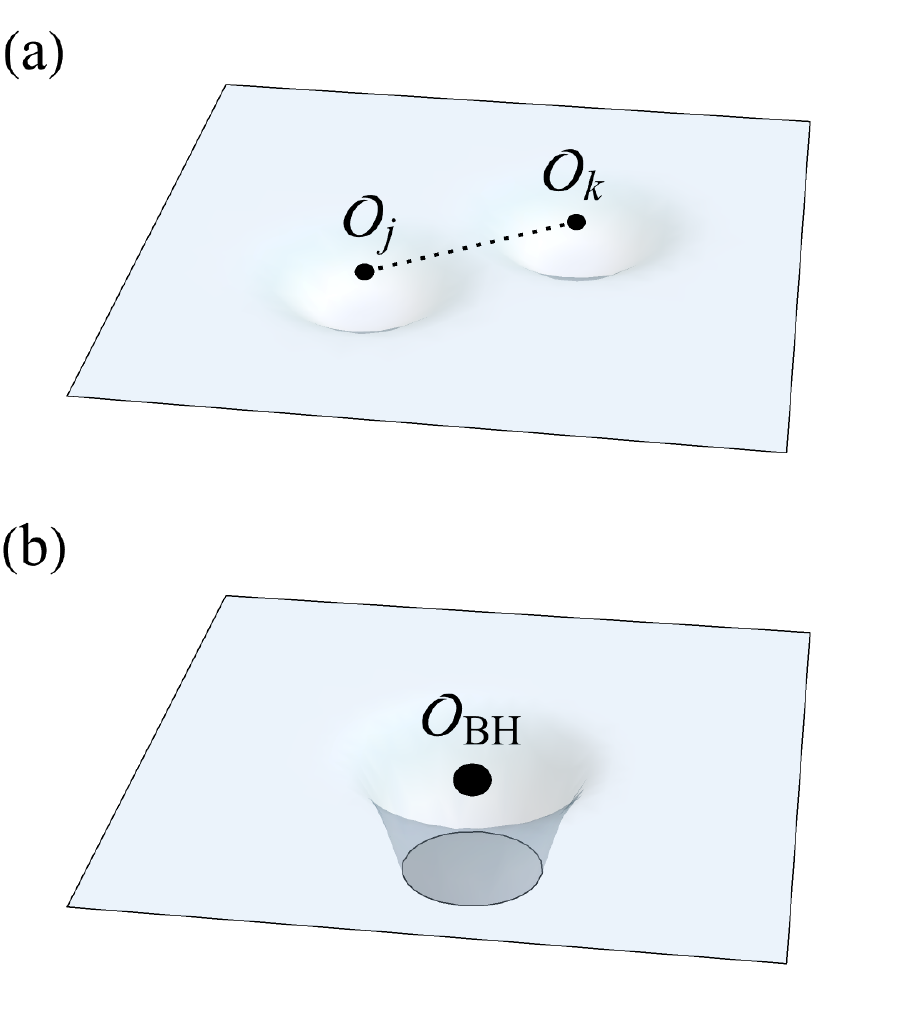}
    \caption{Modifications to the Hilbert space of effective field theory (EFT) by quantum gravity.
    (a) Two EFT operators $\mathcal{O}_j$ and $\mathcal{O}_k$ will, as a result of gravity,
    lose exact commutativity even on the perturbative level, introducing small non-locality (dashed line).
    (b) Any operator $\mathcal{O}_\text{BH}$ producing a black hole state restricts the semi-classical EFT background, leading to a Hilbert space truncation.
    }
    \label{fig:hilbert_trunc}
\end{figure}

Intriguingly, quantum gravity is widely understood to exhibit a qualitatively similar degree of freedom counting problem~\cite{Bekenstein1,Bekenstein:1993dz,Bousso:1999xy,tHooft:1993dmi,Susskind:1994vu}. While the number of degrees of freedom in a local quantum field theory grows volumetrically, in quantum gravity we expect entropy to grow with surface area as exemplified by black hole entropy \cite{Bekenstein:1973ur,Hawking:1974sw} and holography \cite{tHooft:1993dmi,Susskind:1994vu,Maldacena:1997re}.
This problem not only applies to the interior of black holes \cite{akers2022black} but also to the vacuum when the spatial curvature is non-negative. Even in the context of AdS/CFT, the same is expected for a theory with sub-AdS locality. Although it is widely accepted in holography that the Hilbert space of quantum gravity should indeed be smaller than expected from the effective field theory (EFT), an explicit resolution of this degree of freedom counting problem remains open. In particular, this would require a mechanism that on the one hand reproduces the notion of locality in the EFT and on the other accommodates these degrees of freedom into a much smaller quantum gravity Hilbert space. 

In this work, we stress the importance of approximate orthogonality in the context of quantum gravity and cosmology, e.g. \cite{Balasubramanian:2022gmo,Balasubramanian:2022lnw,akers2021quantum,akers2022black,Marolf:2020xie}, and revisit this idea from a different perspective.
We first draw upon an analogy between the Hilbert space dimension verification problem in quantum information and the degree of freedom counting problem in quantum gravity. Furthermore, we propose that the illusion of a bigger Hilbert space in local EFT can be reconciled with the smaller Hilbert space of quantum gravity if the ``qubits'' that make up the EFT are approximate qubits that overlap. In the field-theoretic language, it means that for spacelike separated observables $O(x),O(y)$, $[O(x),O(y)]|\psi\rangle\ne 0$, such that microcausality can be broken for some states $|\psi\rangle$. 
Though such non-locality may seem unphysical, it is well-established that the algebra of local observables for a spacetime region is no longer well-defined in the presence of gravity \cite{Marolf:2020xie,Marolf_2015}, as we visualize in Fig.\ \ref{fig:hilbert_trunc}. For example, \cite{Papadodimas:2013jku,Donnelly:2011hn,Donnelly:2016rvo} showed that gravitational dressing causes apparent local degrees of freedom to overlap even at the perturbative level. Furthermore, a ``verification protocol'' of the quantum degrees of freedom promised by a local EFT, such as creating massive states with low energy density, fails due to black hole formation.

To further sharpen this proposal, we introduce a novel framework for understanding approximate, overlapping qubits which form an approximate algebra of observables for a local region in spacetime using general non-isometric maps. 
These ideas follow a similar spirit as \cite{akers2022black,chao2017overlapping}, but combine both perspectives and build upon their merits. 
We put forward a generalization of the mode transformation used in the original overlapping qubit construction \cite{chao2017overlapping} to express non-isometricity in the language of commutators.
On the one hand, this introduces more versatile and intuitive methods to overlap qubits; on the other hand, it endows richer local structures absent in \cite{akers2022black}. We show by example that by endowing more structures to a non-isometric map than their Haar-random counterparts, more physically relevant scenarios can be constructed where overlap can depend on distance and that the familiar physics such as locality, low energy dynamics, and unitarity can be approximately preserved. In addition, previously state-independent quantities such as microcausality, enforced by vanishing commutators for spacelike separated observables, become state-dependent. We claim that the overlap between such approximate qubits in some constructions may in part be identified with the weak non-locality expected in gauge theory, such as the non-local effects introduced by gravitational dressing. We further identify the discernible departure of the spoofed dynamics from the local EFT with the effects of gravity. As a consequence, the state-dependent commutation relations in the spoofed system are analogous to the effects from gravitational backreactions that shift the underlying spacetime geometry. When the spoofing breaks down and the state-dependent contribution term reaches $O(1)$, this corresponds to the breakdown of the local EFT. This is analogous to strong  gravitational effects dominating and the background causal structure being significantly altered, e.g.\ by the formation of black holes when accessing the massive states in the EFT.

Beyond the above statements which hold in generality, we also apply the formalism to a tensor network toy model of de Sitter spacetime \cite{bao2017sitter}. This spacetime, in addition to being physically relevant for our own expanding universe, presents a case where the tension between EFT and quantum gravity degrees of freedom counting is at its strongest. 
For a system with de Sitter entropy $S_{dS}$, we show that it is possible to construct a strongly complementarian picture following Susskind \cite{susskind2021sitter,Susskind:2021esx} where a seemingly global de Sitter space can arise from a Hilbert space dimension of only $O(\exp(S_{dS}))$. We argue that for any local observer living in the de Sitter universe, the EFT description of global de Sitter may be approximately preserved for up to time $T\sim S_{dS}$ in units of Hubble time. This can be an exceedingly long time for a de Sitter universe with our current cosmological constant, where $S_{dS}\approx 10^{122}$. However, such an effective description is necessarily unstable beyond time $T$, where a transition to a new effective background geometry for the EFT must occur, consistent with the expectation of \cite{Dubovsky:2008rf}. 
Complementary to this line of thought, it has been proposed that the Hilbert space of quantum gravity states that remain asymptotically de Sitter is finite-dimensional \cite{Chakraborty:2023los}. 
We also construct an alternative, weakly complementarian picture, where the number of degrees of freedom would grow linearly instead of exponentially with time and where the notion of local physics does not break down at late times. Compared to past constructions \cite{chao2017overlapping,akers2021quantum} based on random maps, the tensor network examples also demonstrate a connection between internal structures of the non-isometric maps and the emergence of spatial locality among overlapping qubits. We then discuss a few observations on complexity, tensor networks, and von Neumann algebras.

Our contributions are organized as follows. In Sec~\ref{sec:OQ}, we provide a new construction of approximate overlapping qubits based on non-isometric maps, which is different from both Refs.\ \cite{chao2017overlapping} and \cite{akers2021quantum,akers2022black}. We then discuss the consequences of spoofing real qubits with overlapping ones. 
In Sec~\ref{sec:deSitter}, we describe how an expanding spacetime  can be spoofed by overlapping qubits using two different MERA tensor network toy models. In the first model, the fundamental Hilbert space is finite, though the volume of time slices in the EFT description (and hence, the number of overlapping qubits) grows exponentially in time. This effective description is then only valid for finite time and the overlap is distance-dependent. 
A similar distance dependence can be observed in the second de Sitter model, which only describes the EFT in the local patch, leading to a scenario in which spoofing can persist indefinitely. 
Further details of the proof and discussions are given in the Appendices.

\section{Physics of approximate overlapping qubits}\label{sec:OQ}
Let $n$ exact fundamental qubits -- or more generally, qudits -- live in a Hilbert space $\mathcal{H}_n$. It is possible to fit $N>n$ approximate \emph{overlapping qubits} into this Hilbert space, over which a quasi-local effective theory that is approximately unitary can be defined.
In what follows, we generalize the explicit proposal for overlapping qubits of Ref.\ \cite{chao2017overlapping} twofold: First, we do not require each individual qubit algebra to be exact, but allow small overlaps between the local (approximate) Pauli operators as well. 
Second, rather than merely considering mode transformations on $\mathcal{H}_n$ to construct the overlapping qubit algebra, we consider a more general surjective linear map $V:\mathcal{H}_N\rightarrow \mathcal{H}_n$. This map is necessarily non-isometric for $N>n$.
Let $Q\in L(\mathcal{H}_N)$ and $Q_p=VQV^{\dagger}\in L(\mathcal{H}_{n})$ be bounded linear operators over their respective Hilbert spaces and $|\psi_p\rangle=V|\psi\rangle$. We say that $Q_p$ acts on an approximate overlapping qubit if $Q$ is supported on a qubit subsystem in $\mathcal{H}_N$. The action of $\prod_i Q^{(i)}$ in the effective theory can be \emph{spoofed} with respect to $|\psi\rangle$ if $\langle \psi_p|\prod_i^MQ_p^{(i)}|\psi_p\rangle\approx \langle\psi|\prod_i^M Q^{(i)}|\psi\rangle$.
It is easy to see that for suitable maps $V$, operators, and states $\ket\psi$ (potentially restricted to a physical subspace), unitarity and algebraic structure are also approximately preserved, with $\langle\psi|U^{\dagger}U|\psi\rangle\approx \langle \psi_p|U^{\dagger}_p U_p|\psi_p\rangle$ and $\Vert [Q,Q']|\psi\rangle\Vert  \approx \Vert [Q_p,Q'_p]|\psi_p\rangle\Vert $, and similarly for the anti-commutation relations. More generally, one can identify such state-operator pairs given a form of $V$.
\begin{figure}[t]
    \centering\begin{tikzpicture}[scale=0.88]
        \draw[red, thick, fill=red!20] (0,0.5) circle (0.5cm);
        \node[color=red,anchor=south] at (0,1.0) {$X,Z$};
        \draw[blue, thick, fill=blue!20] (1.5,0.5) circle (0.5cm);
        \node[color=blue,anchor=south] at (1.5,1.0) {$X,Z$};
        \draw[purple, thick, fill=purple!20] (3,0.5) circle (0.5cm);
        \node[color=purple,anchor=south] at (3,1.0) {$X,Z$};
        \draw[green, thick, fill=green!20] (0.75,-0.5) circle (0.5cm);
        \node[color=green,anchor=north] at (0.75,-1.0) {$X,Z$};
        \draw[orange, thick, fill=orange!20] (2.25,-0.5) circle (0.5cm);
        \node[color=orange,anchor=north] at (2.25,-1.0) {$X,Z$};
    \draw[thick,-latex] (4,0)--(5,0) node [midway, above] {\textbf{V}};
    \fill[rounded corners=0.5mm,red!20] (6,0.3) \irregularcircle{0.5cm}{0.5mm};
    \fill[rounded corners=0.5mm,blue!20] (6.5,0) \irregularcircle{0.5cm}{0.5mm};
    \fill[rounded corners=0.5mm,purple!20] (7.25,0.25) \irregularcircle{0.5cm}{0.5mm};
    \fill[rounded corners=0.5mm,green!20] (5.8,-0.5) \irregularcircle{0.5cm}{0.5mm};
    \fill[rounded corners=0.5mm,orange!20] (7.1,-0.5) \irregularcircle{0.5cm}{0.5mm}; 
    \draw[rounded corners=0.5mm,red,thick] (6,0.3) \irregularcircle{0.5cm}{0.5mm};
    \draw[rounded corners=0.5mm,blue,thick] (6.5,0) \irregularcircle{0.5cm}{0.5mm};
    \draw[rounded corners=0.5mm,purple,thick] (7.25,0.25) \irregularcircle{0.5cm}{0.5mm};
    \draw[rounded corners=0.5mm,green,thick] (5.8,-0.5) \irregularcircle{0.5cm}{0.5mm};
    \draw[rounded corners=0.5mm,orange,thick] (7.1,-0.5) \irregularcircle{0.5cm}{0.5mm};
    \node[red,anchor=south east] at (6,0.8) {$X_{p},Z_{p}$};
    \node[blue,anchor=south] at (6.5,0.8) {$X_{p},Z_{p}$};
    \node[purple,anchor=south west] at (7.25,0.75) {$X_{p},Z_{p}$};
    \node[green,anchor=north] at (5.8,-1.1) {$X_{p},Z_{p}$};
    \node[orange,anchor=north] at (7.1,-1.1) {$X_{p},Z_{p}$};
    \draw[thick] (-0.75,1.6) rectangle (3.75,-1.6);
    \node at (4.15,1.2) {$\mathcal{H}_{N}$};
    \draw[thick] (5.2,-1.1) rectangle (7.9,0.85);
    \node[anchor=west] at (7.85,0.5) {$\mathcal{H}_{n}$};
    \end{tikzpicture}
    \caption{The non-isometric map $V$ maps nominally exact qubits (circles) in $\mathcal{H}_N$ onto approximate overlapping qubits (jagged circles) in a lower-dim.\ $\mathcal{H}_n$ where $\{X_p,Z_p\}\approx 0$. }
\end{figure}
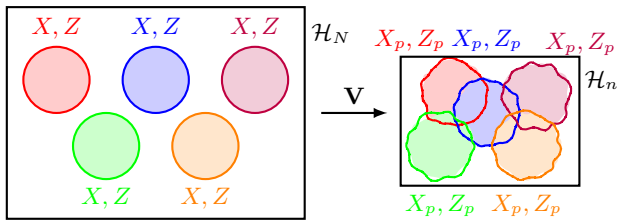
\begin{remark}\label{prop}\textup{(Approximate homomorphism property.)}

Let $P=V^{\dagger}V$ be a truncation map. The set of processes $\mathcal{S}=\{\prod_i^{\leq M} Q^{(i)}\}$ is well-spoofed, i.e., $\langle \psi_p|\prod_i^{\leq M} Q_p^{(i)}|\psi_p\rangle\approx \langle \psi|\prod_i^{\leq M} Q^{(i)}|\psi\rangle$, if $\forall S\in \mathcal{S}, \Vert P S|\psi\rangle- S|\psi\rangle\Vert \approx 0$.
\end{remark}
In other words, a sufficient condition for identifying well-spoofed processes is if the string of operators $Q^{(i)}$ do not take $|\psi\rangle$ out of the subspace that is approximately preserved by $V$. In general, given a $V$ (e.g.\ acting on the Hilbert space of black holes \cite{akers2022black}) one can also identify other well-spoofed processes by solving a set of constraint satisfaction problems (see Appendix~\ref{app:proof} for details).
For any such $V$, one can perform a singular value decomposition and write $P=\sum_k \lambda_k|\psi_k\rangle\langle \psi_k|$ with orthonormal set of states $\{|\psi_i\rangle\in\mathcal{H}_N\}$ (note that $P$ is not a projection in general). The processes preserved by spoofing naturally depends on the form of $V$.  %
For example, suppose $V$ is a global truncation onto the low energy subspace of some local Hamiltonian, e.g. that of a local QFT, such that for some global energy cut off $\Lambda$, $\lambda_{k\leq\Lambda}=1$, then the processes $\prod_i Q^{(i)}$ that can be effectively spoofed are ``low energy processes'' which take $|\psi\rangle$ to states that have negligible support over the high energy subspace with $k>\Lambda$ \cite{Arad:2014znf,Crosson}. Such is expected for a local theory and associated operators as a consequence of UV-IR decoupling. A similar conclusion holds for our later MERA tensor network construction, as it prepares states with power-law correlations related to conformal field theories.
If, on the other hand, $V$ is proportional to a Haar-random projection as in \cite{akers2021quantum,akers2022black}, then with high probability, these subspace preserving processes are operators with subexponential complexity if $|\psi\rangle\not\in \operatorname{ker}(V)$.
Returning to general $V$s, we note that these approximate qubit operators also overlap. This overlap quantifies how well locality is preserved by the spoofing. For any $Q,Q'$, 
$$[Q_p,Q_p']|\psi_p\rangle = V[Q,Q']V^{\dagger}|\psi_p\rangle + T^{[Q,Q']}_p|\psi_p\rangle,$$ 
where $T^{[Q,Q']} = QTQ'-Q'TQ$ and $T=V^{\dagger}V-I$. If $[Q,Q']=0$, then the projected operators indeed overlap by an amount given by $\Vert T_p^{[Q,Q']}|\psi_p\rangle\Vert $. Depending on the spectral property $\spec(T_p^{[Q,Q']})$, we can immediately deduce: 1) If $\exists \lambda\in \spec(T_p^{[Q,Q']})$ such that $\lambda<\epsilon$, then there exist states $|\psi_p\rangle$ for which the locality is approximately preserved, i.e., $\Vert [Q_p,Q'_p]|\psi_p\rangle\Vert  =O(\epsilon)$. For instance, this is the case if $Q,Q'$ acting on $|\psi\rangle$ are well-spoofed processes we examined above. 2) The value of the commutation relation is state-dependent if $\spec(T_p^{[Q,Q']})$ is non-flat.  We find that both are the case for the $V$s we consider in this work, where the spectrum is non-flat with many of its eigenvalues are concentrated near zero (Fig.~\ref{fig:ising-dep}a). For example, we show that if $V$ is a low energy truncation of a 1d Ising CFT, then commutators of local observables with respect to low-energy states are relatively well-preserved. Interestingly, the states that give rise to larger microcausality violation also correlate with higher energies (Fig.~\ref{fig:ising-dep}b). See Appendix \ref{app:spec} for details. 

\begin{figure}
    \centering
    \includegraphics[width=\linewidth]{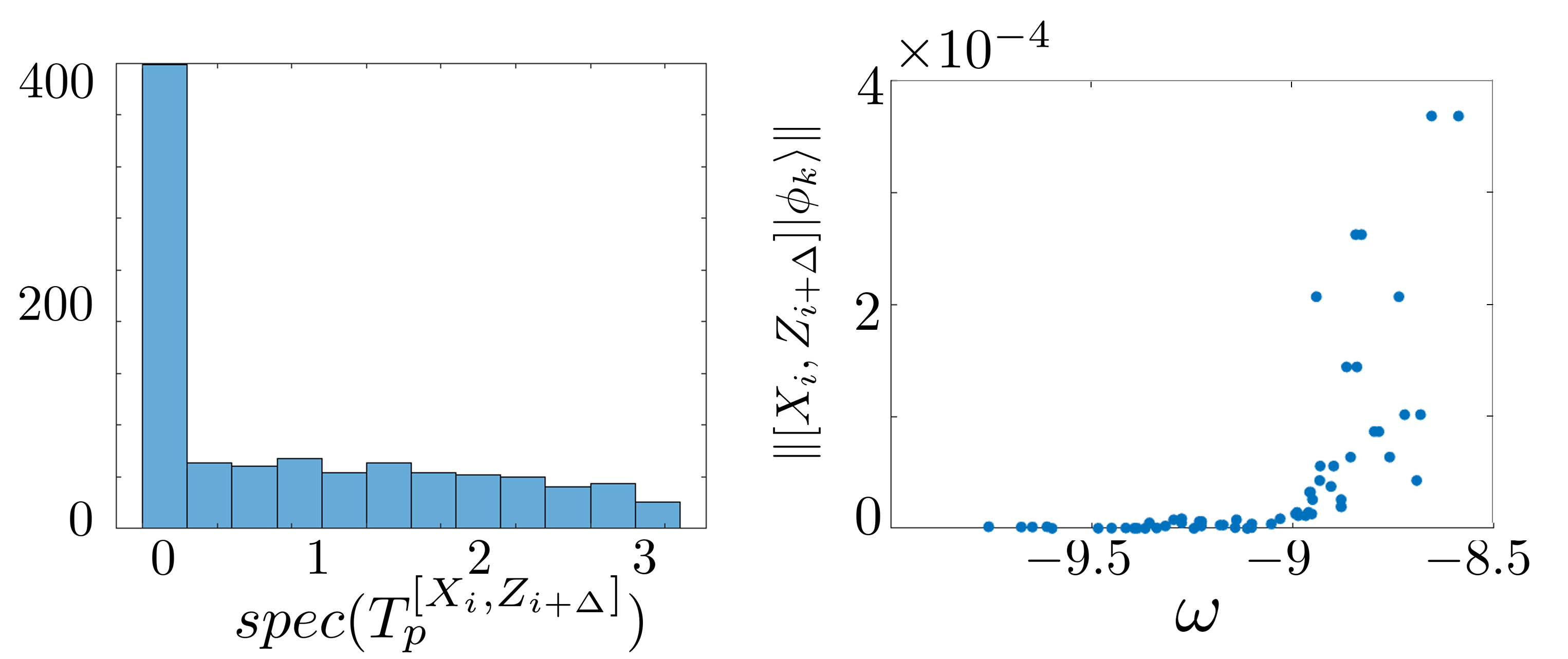}
    \caption{For $V$ being a low energy truncation of a 1d critical Ising CFT, (a) shows the eigenvalue distribution of $T_p^{[X_{i},Z_{i+\Delta}]}$ as a histogram, where $X_{i},Z_{i+\Delta}$ are local Pauli operators separated by a fixed spatial distance $\Delta$. (b) shows a correlation between the amount of microcausality violation and the ``energy'' $\omega$ of each state $|\phi_{k}\rangle$ labelled by different $k$s, such that more energetic states generally incur greater violations, consistent with our intuition. The energy $\omega$ is defined by a truncated critical Ising Hamiltonian. See Appendix \ref{app:tij} for detailed definitions.
    }
    \label{fig:ising-dep}
\end{figure}

These observations carry physical significance in the context of quantum gravity. In perturbative quantum gravity, if $Q,Q'$ are gravitationally dressed operators \cite{Donnelly:2016rvo} that are spacelike separated, then the small but non-vanishing commutator of $[Q_p,Q_p']|\psi_p\rangle$ is consistent with the gauge invariant operator being weakly non-local by turning on gravity. Formally, $T^{[Q,Q']}_p$ plays the role of the dressing, where it is generally a non-local operator straddling the local operators $Q,Q'$. On the other hand, changes to the state $|\psi\rangle$ would generally signal a different configuration for the stress-energy, which through gravity, leads to a different background spacetime geometry. In this sense, we should also expect the causal structure defined by the commutators to be state-dependent. Hence, we identify the departure of the actual physics obtained via spoofing from the local EFT on a fixed background with the effect of gravity or potentially new physics beyond local EFT-based predictions in the Standard Model or $\Lambda$CDM. 

Finally,  as the fundamental processes cannot spoof everything promised by the EFT in a much larger Hilbert space, there are regimes for which $\lambda$ is large and $T_p^{[Q,Q']}|\psi_p\rangle$ dominates. Such is also the premise behind the verification protocol of Ref.\ \cite{chao2017overlapping}. 
Then the effective description with $[Q,Q']|\psi\rangle=0$ is not even approximately true. This is similar to the scenario where the effect of quantum gravity dominates and the local EFT is expected to break down. From Remark \ref{prop}, we see that this may be because the effective processes now have large support over the subspaces that are being cut off. In the global energy truncation, it corresponds to considering $|\psi\rangle\in \ker(V)$, which is also known as the null space of the non-isometric map. This may happen in quantum gravity, for instance, when a large black hole has evolved long past the Page time. We show that this breakdown also occurs for EFT states at late times for the de Sitter MERA model we introduce below. 

\section{Overlapping qubits in de Sitter}
\label{sec:deSitter}

\begin{figure}[t]
    \centering
    \includegraphics[width=0.35\textwidth]{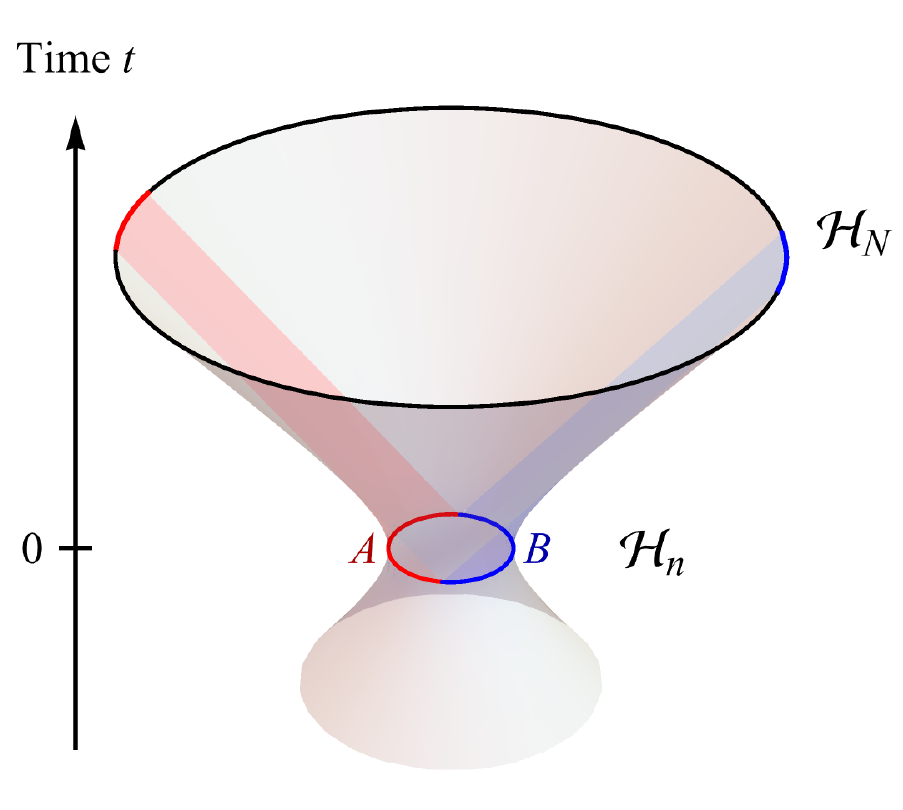}
    \caption{Global De Sitter spacetime as a hyperboloid in a higher-dimensional Minkowski embedding. The volume of time-slices increases exponentially with global time $t$, with a minimum effective Hilbert space $\mathcal{H}_n$ at $t=0$. We propose a projection of effective Hilbert spaces $\mathcal{H}_N$ at $t>0$ into the ``fundamental'' Hilbert $\mathcal{H}_n$.
    Two causal patches with constant volume are shaded in red and blue, emanating from two complementary regions $A$ and $B$ at $t=0$.    
    }
    \label{fig:ds_slices}
\end{figure}

As we have seen, overlapping qubits resulting from a non-isometric truncation of EFT Hilbert spaces produce effects that are in principle compatible with those of quantum gravity effects. However, we did not yet specify the form of such a non-isometric map or the physical setting in which it may occur. As a specific example to explore our idea, we now consider the case of de Sitter (dS) spacetime, a model of an expanding universe. A long mystery of semiclassical dS spacetime has been the relationship between the EFT Hilbert spaces living on different time-slices: It appears that as the spatial volume increases exponentially with time $t$, so does the size of EFT Hilbert space. It has therefore been suggested that the usual unitary evolution rules of quantum mechanics have to be replaced by \emph{isometric} evolution in the dS case \cite{Cotler:2022weg,Cotler:2023eza}. In fact, such an isometric map appears to become neccessary in any dynamical spacetime with a UV cutoff \cite{Hohn:2014uvt}. A potential resolution to this problem is that while the EFT Hilbert space may be growing, the \emph{fundamental} Hilbert space may remain constant, an idea similar to which has recently been applied to black hole evaporation \cite{akers2022black}. 
Given the isometric evolution map $V$ of the dS setting, a non-isometric map to a smaller Hilbert space is readily available: It is given by $V^\dagger$, inducing a truncation $V V^\dagger$ on the EFT Hilbert space. This leaves the choice of an initial time-slice with which the fundamental Hilbert space is associated. Globally, there is only one coordinate-independent choice to make, the time-slice of minimal volume (at $t=0$ in the standard coordinates), as we show in Fig.\ \ref{fig:ds_slices}.\footnote{In other spacetimes with time-dependent expansion or contraction, there may be several extremal-volume Cauchy slices, complicating the choice of the ``fundamental'' slice.}  
As we show below, one can extend this proposal to time-slice subregions as well, where again a fundamental Hilbert space may be associated with a minimal surface.

\begin{figure}[t]
    \centering
    \includegraphics[width=0.48\textwidth]{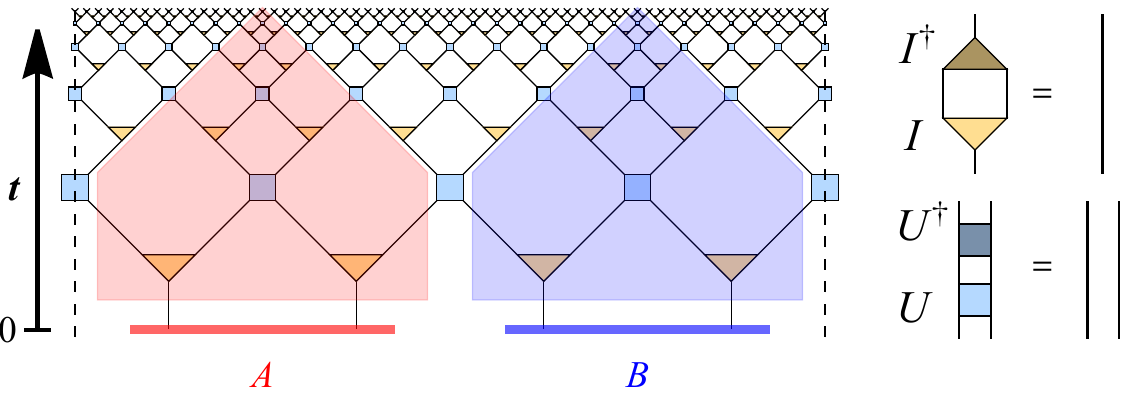}
    \caption{Discretization oft two-dimensional de Sitter spacetime dS$_2$ with the MERA tensor network consisting of ``disentangling'' unitaries $U$ (square) and isometric tensors $I$ (triangles) whose properties are shown on the right. The left and right edges of the the tensor network (dashed lines) are periodically identified. The ``entanglement renormalization'' direction of the MERA is identified with global dS time $t$. 
    The two complementary static patches from Fig.\ \ref{fig:ds_slices} are shown here as well, again shaded in red and blue.
    }
    \label{fig:ds_wedges}
\end{figure}

To explore the concrete behavior of the non-isometric maps in dS in a finite setting, we will be considering a discretization of dS spacetime in terms of the MERA tensor network, which possesses the same causal structure. We begin by reviewing the properties of this tensor network model.

\subsection*{Review of de Sitter MERA}\label{rev:MERA}

The \emph{multi-scale entanglement renormalization ansatz} (MERA) is a tensor network ansatz designed to produce (ground) states of critical Hamiltonians \cite{Vidal:2008zz}, used as lattice models of conformal field theories \cite{Pfeifer:2009criticalMERA}. It is most commonly used for $1{+}1$-dimensional theories, where the tensor network takes the form of a planar branching tree with interconnections on each layer. Even though this geometry breaks translation invariance, the MERA can be used to efficiently simulate translation-invariant models \cite{Evenbly:2007hxg}. 
Along the branching direction of the network a flow of \emph{entanglement renormalization} can be identified \cite{Pfeifer:2009criticalMERA, Vidal.ERintro}. As this resembles the radial direction of an AdS time-slice in AdS/CFT, it was proposed that the MERA is a discrete model of AdS holography \cite{Swingle:2009bg}. However, it was later argued that the tensor network geometry of the MERA does not directly correspond to the spatial geometry of an AdS time-slice \cite{Bao:2015uaa, czech2016tensor}.
Rather than a spatial coordinate in a negatively curved space-time, another interpretation of this branching direction is that of a time coordinate in positively curved de Sitter space-time \cite{Beny:2011vh,czech2016tensor,SinaiKunkolienkar:2016lgg,bao2017sitter,osborne}. In this picture, each MERA layer corresponds to a time-slice of an expanding universe, with each tensor representing a piece of spacetime with spatial and temporal extent of the order of the Hubble radius and Hubble (doubling) time, respectively.
The tensor network structure of the MERA, using the standard MERA constraints on unitary disentanglers and isometric branching tensors, reproduces the causal structure of the $1+1d$ global dS spacetime. In particular, for a $k$-branching MERA there exists a tensor network notion of a \emph{static patch}, the part of the universe visible to a stationary observer: It is given by a wedge comprised of $k$ isometries and $k-1$ disentanglers at each timeslice, as shown in Fig.\ \ref{fig:ds_wedges} for $k=2$. As in continuum de Sitter, one can divide the entire space at $t=0$ into two halves $A$ and $B$, each centered around a ``pode'' and an ``antipode'', i.e., two stationary observers. Keeping the causally accessible region for each observer by tracing along the timelike paths in the network yields two regions whose (proper) volumes are fixed under exponential expansion with $t$; the resulting spacetime volumes (red and blue regions in Fig.\ \ref{fig:ds_wedges}) are the two static patches of the initial observers.

In a scale-invariant MERA, the state inside the static patch also tends to a fixed point, yielding an expected outcome from the ``cosmic no-hair theorem'' \cite{bao2017sitter}. For each spatial slice, the de Sitter entropy in this model is upper bounded by the edge cuts that separate the interior of the patch from the exterior. As each tensor is of Hubble scale, we have the bond dimension $\chi\sim O(\exp(e^{S_{dS}}))$ for the MERA tensor network.

With the correct choice of tensors, the MERA also produces the ground state of many critical lattice models with a CFT continuum limit at $t \to \infty$ \cite{Evenbly:2007hxg,Pfeifer:2009criticalMERA}. In the de Sitter picture, the sites associated with that state are located at future infinity $\mathcal{I}^+$. As the symmetry group $SO(d,1)$ is both obeyed by empty de Sitter spacetime as well as a Euclidean CFT, it is tempting to identify this asymptotic CFT as a ``boundary dual'' of dS gravity. However, unlike AdS/CFT, the boundary lacks a time-like coordinate, precluding a dynamical dictionary, and the ``duality map'' between de Sitter space-time and $\mathcal{I}^+$ is simply the causal evolution of the universe. It is worth noting that other interpretations of the MERA also exist. For example, when relating MERA to path integral geometries, it was argued that MERA network describes a light-like hypersurface in AdS spacetime \cite{Milsted}. 
As we make no connection with path integral geometries or the AdS/CFT correspondence, these interpretations are irrelevant for our purposes, where the salient feature of the MERA is its causal structure which matches discretized de Sitter spacetime.
We also hasten to point out that the critical state a MERA can be tuned to produce at $\mathcal{I}^+$ has no special meaning here and should not be confused with the usual emergence of a semi-classical bulk in AdS/CFT.

Using the MERA as a coarse-graining description of de Sitter spacetime, one immediately realizes the proposed isometric time-evolution from a time-slice at earlier time to one at later time \cite{Cotler:2023eza}, in that the state produced by the MERA lives in a nominal Hilbert space whose size doubles at every layer. However, using suitable cuts through the MERA we can also interpret it as a map from this nominal, rapidly expanding Hilbert space to a much smaller ``fundamental'' Hilbert space. In this concrete tensor network setting, we can explore how such maps constrain the space of operators or states in the nominal Hilbert space. We now propose two such MERA-based map: A ``global'' proposal in which the fundamental degrees of freedom stay constant with time $t$, and a ``local'' proposal where their growth is merely linear, rather than exponential.

Although we mostly make general statements about the MERA, it is also helpful at times to consider specific examples where disentanglers and isometries are fixed to particular values. For benchmarking purposes, we consider three explicit instances. For the first instance, we instantiate tensors of bond dimension  $\chi=2$ that optimize the ground state of the critical transverse field Ising model, whose continuum limit is the Ising CFT. The details of this construction are explained in Appendix \ref{app:ising}. We also consider a model with Haar-random unitaries selected as disentanglers and isometries with different choices of  $\chi$. These examples are studied numerically. Finally, we consider a more general construction at large $\chi$ based on analytical arguments. Generally, we expect the large $\chi$ limit to be necessary for realistic models of de Sitter spacetime, so that each static patch is described by sufficiently many parameters to resolve details below the Hubble scale.

\begin{figure*}[ht]
    \centering
    \includegraphics[width=0.42\textwidth]{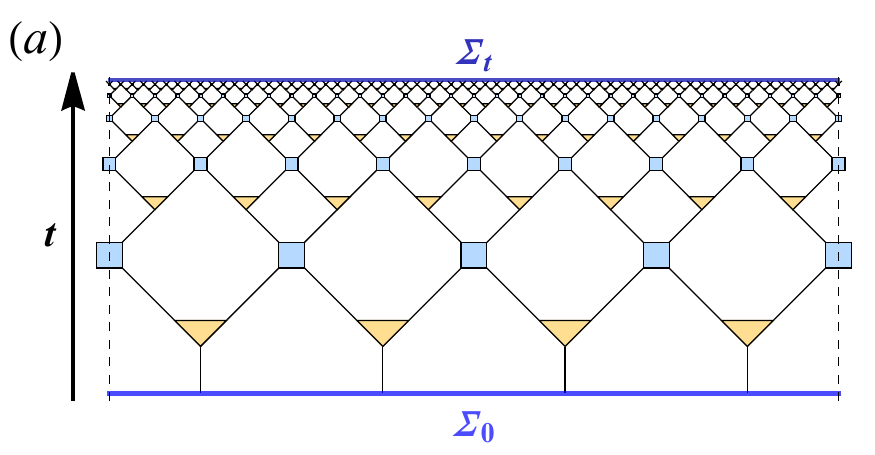}
    \hspace{0.5cm}
    \includegraphics[width=0.42\textwidth]{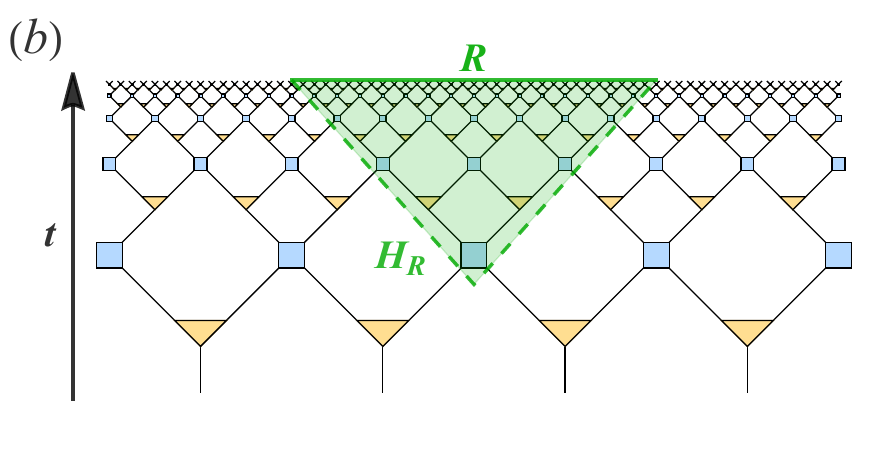}
    \caption{Global and local non-isometric maps in a MERA discretization of de Sitter spacetime. 
    (a) In the global proposal, we consider a map $V_\text{global}$ from the degrees of freedom on a time-slice $\Sigma_t$ at global time $t$ to an initial time-slice $\Sigma_0$. While the Hilbert space dimension on $\Sigma_t$ grows exponentially with $t$, it is constant on $\Sigma_0$. The left and right end of the diagram (dashed lines) are identified.  
    (b) The alternative local proposal involves a map $V_\text{local}$ from any subregion $R$ (of a time-slice $\Sigma_t$) to the (nearly) light-like horizon $H_R$ of its past domain of dependence, with $H_R$ having exponentially smaller Hilbert space dimension.
    }
    \label{fig:ds_proposals_mera}
\end{figure*}
\vspace*{-3mm}

\subsection*{Global dS MERA}\label{sec:global}
Global dS evolution, with the apparent Hilbert space degrees of freedom doubling every Hubble time, can be naturally discretized in terms of the MERA \cite{bao2017sitter}. 
From the finiteness of de Sitter entropy $S_{\rm dS}$, complementarity, and de Sitter holography, it has been suggested that the dimension of the quantum gravity Hilbert space for de Sitter spacetime is finite \cite{Witten:2001kn,Bousso:2000nf,Parikh:2004wh,Dyson:2002pf,Banks:2000fe} and given by $O(\exp(S_{\rm dS}))$, which is proportional to the de Sitter horizon area or the area of the spacelike extremal surface at $t=0$. Refs.\ \cite{Cotler:2022weg,Cotler:2023eza} proposed that this Hilbert space is isomorphic to a subspace of states that do not lead to big crunch cosmologies . On the surface, this finite-dimensional description seems to be at odds with the semiclassical picture that has exponential growth in the spacetime volume. It is also unclear how the expanding spacetime should emerge from a finite dimensional Hilbert space.
Here we suggest a possible construction by describing the expanding global dS space with an increase of overlapping qubits, whose fundamental Hilbert space has dimension $O(\exp(S_{\rm dS}))$.  As a consequence, the apparent exponential expansion can last for a long time before the notion of locality breaks down. 

To identify the overlapping qubits, we build a non-isometric map using the MERA. Let $\mathcal{H}_t\cong\mathcal{H}_{N_t}$ be the apparent Hilbert space associated with the global timeslice $\Sigma_t$ at time $t$ and $\mathcal{H}_{t=0}\cong\mathcal{H}_n$ be the fundamental degrees of freedom tied to the dS quantum gravity with $N_t\geq  n$, where $\dim\mathcal{H}_{t=0}\sim\exp(S_{dS})$. We posit that de Sitter quantum gravity is governed by some dynamical processes over $\mathcal{H}_n$, such that the effective description up to some time $T$ can be approximated by a sequence of non-isometric maps $V_t:\mathcal{H}_{N_t}\rightarrow \mathcal{H}_n$, each of which is given by a MERA cut off at time $t$ (Figure~\ref{fig:ds_proposals_mera}a). Using $V_t$, we can identify the algebra of observables (e.g.\ the ones supported on each site) in $\mathcal{H}_t$ with an approximate algebra in $L(\mathcal{H}_n)$ using our overlapping qubit construction.
This effectively compresses $O(\exp(t))$ number of qubits into an only constant number of qubits at the expense of introducing overlaps. Here let us focus on MERAs that are scale- and translationally invariant, i.e., the disentanglers (and isometries) at each layer are identical. By fixing the individual tensors, an explicit mapping $V_t$ can be constructed. 
As an isometric map, $V_t^{\dagger}$ identifies a subspace $\mathcal{C}\subset \mathcal{H}_t$. We can conclude from Remark~\ref{prop} that operations preserving the subspace $\mathcal{C}\cong\mathcal{H}_{t=0}$ are well-approximated, thus recovering properties like the cosmic no-hair theorem in \cite{bao2017sitter} or those supported in \cite{Cotler:2023eza}. Here $\mathcal{C}$ is a space of states with power-law correlations which one can obtain by only altering the initial conditions at $t=0$. Different from random maps used in \cite{akers2021quantum,akers2022black} and the construction by \cite{chao2017overlapping} based on the Johnson-Lindenstrauss (JL) lemma, the locality of MERA overlaps the qubits by an amount that depends on the proper distance. 
\begin{proposition}
    Let $O^{(i)},O^{(j)}$ be local operators supported on qubits $i$ and $j$ respectively on $\Sigma_t$ separated by proper distance $|i-j|$ in units of Hubble radius in the MERA, then non-isometric maps can be constructed to produce overlap such that for some state $|\psi_p\rangle$ 
\begin{equation}\label{eqn:dsoverlap}
\Vert [O^{(i)}_p,O^{(j)}_p]|\psi_p\rangle\Vert \lesssim \frac{\epsilon(t)}{|i-j|^{\alpha}}
\end{equation}
where the upper bound obeys a power law and that 
$\epsilon(t)\ll 1$ for $t\lesssim O({S}_{\rm dS})$.
\end{proposition}%
We provide a detailed argument in Appendix \ref{app:tij} for certain scale-invariant random MERAs at large bond dimension. Because at each time $t$, the size of $\epsilon$ is suppressed by $\exp(-O(S))$ for this model, the effect of overlaps becomes less visible for universes with smaller cosmological constants, thereby strengthening the notion of locality in the large $\chi$ limit.

Heuristically, we argue that a similar behaviour can be anticipated in the large $\chi$ limit for a less random MERA construction where the tensor network reproduces a CFT ground state. For example, it is expected that for some CFTs, a MERA with large $\chi$ can converge to an exact low energy truncation that preserves the first $\mathrm{poly}(\chi)$ energy eigenstates\footnote{This is not guaranteed, as the validity of MERA has not been demonstrated for CFTs of large central charges and strong interactions.}. Recall that the transition $\langle E|O|\Omega\rangle$ induced by a local operator $O$ between some low-energy state $|\Omega\rangle$ and a high energy state $|E\rangle$ with energy eigenvalue above the cutoff is exponentially suppressed by the energy difference between the two states for local Hamiltonians \cite{Arad:2014znf,Crosson}. As a larger $\chi$ leads to a higher energy cutoff, the notion of locality (\ref{eqn:dsoverlap}) becomes sharper as one increases $\chi$ for such critical MERAs also. This is to say that the overlap between spacelike separated observables with respect to sufficiently low energy states vanish in the large $\chi$ limit.
We hasten to point out that because the MERA constructions are coarse-grained and have no sub-Hubble features, the overlap within each causal patch (\ref{eqn:dsoverlap}) does not decay with distance even though $\epsilon$ is small. As a more realistic tensor network model of our universe with sub-Hubble features has not yet been constructed, we will leave this discussion for future work.

\begin{figure*}[ht]
    \centering
    \includegraphics[width=0.9\textwidth]{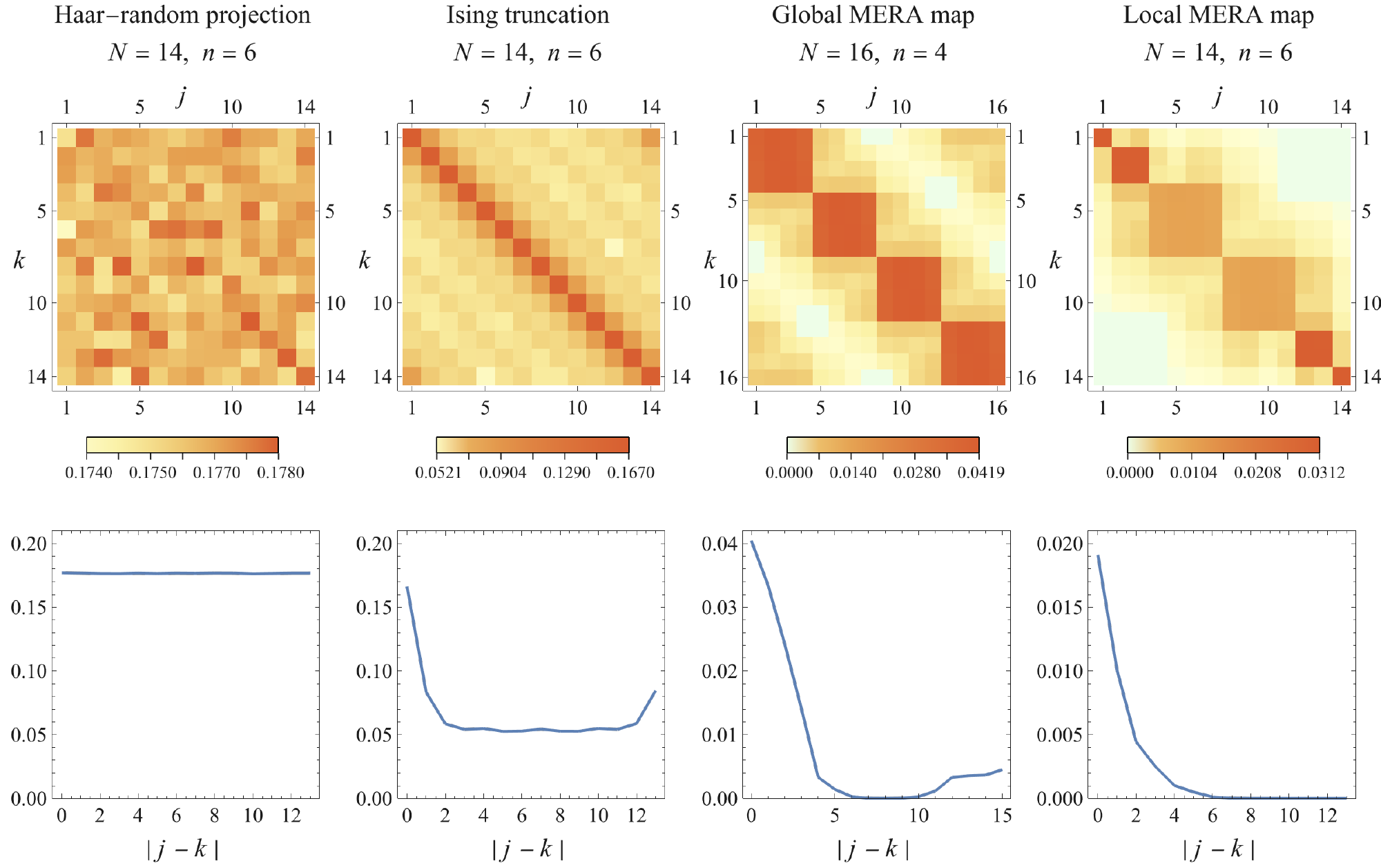}
    \caption{Commutator trace norms $\Vert[X_p^{(j)},Z_p^{(k)}]\Vert_1 / \Vert X_p^{(j)}\Vert_1 \Vert Z_p^{(k)}\Vert_1$ of projected Pauli operators, defined as $X_p^{(j)} = V X^{(j)} V^\dagger$ and $Z_p^{(k)} = V Z^{(k)} V^\dagger$, where $V: \mathcal{H}_N \to \mathcal{H}_n$ is a non-isometric map from an $N$-dimensional to a smaller $n$-dimensional Hilbert space. The examples of $V$ are an Akers-Pennington type Haar-random projection (averaged over 20 samples), a Hilbert space truncation to the low-energy critical Ising model, as well as our global and local MERA map (compare Fig.\ \ref{fig:ds_proposals_mera}) with tensors optimized with respect to the critical Ising model ground state. 
    The dependence of the average commutator norm on distance $|j-k|$ is plotted in the second row, showing a decay for all but the Haar-random projection.
    The Ising truncation leads to a plateau at large distances, but allows for a stronger state-dependent decay as shown in Fig.\ \ref{fig:gscomm_ising} (Appendix~\ref{app:tij}).
    }
    \label{fig:xz_comm1}
\end{figure*}

At finite sizes, we also verify the distance dependence numerically with examples using a MERA optimized for an Ising CFT (Fig.~\ref{fig:xz_comm1}) or using random tensors (Appendix~\ref{app:ising}). Note that some of the MERA overlaps vanish because of discretization as the past-causal cone of some sites do not overlap with a finite-time cut off. As this computes the commutator norm, it puts an upper bound on the amount of overlap one has for any state. 

It is more physically relevant to examine the overlap with respect to a particular state. For example, the commutator is state-dependent where the amount of overlap weakly correlates with the energy of the state with respect to the Ising CFT Hamiltonian above a certain threshold (Fig.~\ref{fig:ising-dep}b).
It is clear that the overlaps with respect to select states are far smaller than the overall norm in Fig.~\ref{fig:xz_comm1}. See also Fig.~\ref{fig:gscomm_ising} in Appendix~\ref{app:tij} for spatial overlap with respect to the ground state.

However, this approximation cannot persist indefinitely as we eventually run out of space to accommodate the overlapping qubits. In units of Hubble time, the approximation is only valid for $T\sim O(S_{dS})$ before $\epsilon\sim O(1)$, a prediction consistent with the scaling found in Ref.\ \cite{Dubovsky:2008rf}. 
In Appendix \ref{app:tij}, we show that such an estimate is reproduced by a MERA model of the dS static patch.
$M$-point functions inside a static patch would hold for a similar time scale up to constant multiplicative factors and logarithmic corrections. 
\begin{proposition}
    Let $Q^{i_k}\in L(\mathcal{H}_{N_t})$ be observables in the static patch. Then for constant $M$, there exist MERA non-isometric maps $V$ such that $M$-point correlations are well-spoofed
    \begin{align*}
        &|\langle\psi_p|Q^{i_M}_p\dots Q^{i_1}_p|\psi_p\rangle- \langle \psi|Q^{i_M}\dots Q^{i_1}|\psi\rangle| \sim O(\epsilon)
    \end{align*}
    where $\epsilon<O(1)$ up to time $T\lesssim S/2M.$
\end{proposition}
A similar time scale for $T$ can also be obtained using a JL-based construction in \cite{chao2017overlapping} which would permit constant overlap for qubits separated by super-Hubble distances. 

\vspace*{-2.5mm}
\subsection*{Local dS MERA}\label{sec:local}
More recent proposals for a holographic interpretation of de Sitter spacetime have suggested that a holographic theory may be encoded on stretched horizons that are nearly light-like \cite{susskind2021sitter,Shaghoulian:2021cef}. Such an approach readily leads to a second construction for the implementation of a non-isometric map that reduces the apparent number of degrees of freedom on time-slices. 
We construct a non-isometric map from spatial subregions $R \subset \Sigma_t$ on a de Sitter time-slice $\Sigma_t$ to the fundamental degrees of freedom located at the stretched horizon $H_R$ of its past domain of dependence.
Assuming such a map to a fundamental Hilbert space on nearly light-like surfaces is well-defined, we can immediately see from the MERA discretization (Fig.\ \ref{fig:ds_proposals_mera}(b)) that it implies an exponential reduction in degrees of freedom. Mirroring the causal structure in the continuum, the wedge $\mathcal{W}_R$ bounded by $\partial \mathcal{W}_R = R \cup H_R$ contains the MERA tensors that are irrelevant to computing expectation values of operators with support only on $R^c$, the complement region to $R$.
In particular, if $R$ is chosen to cover half of the global space, its wedge $\mathcal{W}_R$ corresponds to the ``exterior region'' between two initial static patches in $1{+}1$ dimensions. $H_R$ then overlaps with the horizons of both patches.  Interestingly, the entanglement structure is altered by the compression map $V_{\rm local}$, such that it converts a non-flat continuous spectrum for a state in the exterior region to one of nearly maximally entangled pairs along the stretched horizon. As overlaps are analogous to perturbative gravitational effects, in the infinite dimensional limit this is qualitatively similar to \cite{chandrasekaran2022algebra}, where it was found that gravitational effects convert a type III von Neumann algebra type II$_1$, even though the conversion occurred on different spacetime regions. We further comment on this in Appendix \ref{app:static-patch-ee}.

The behavior of commutators is quite similar between the global and local map: As we find in Fig.\ \ref{fig:xz_comm1}, both show a decay of the trace norm of the commutator $\Vert[X_p^{(j)},Z_p^{(k)}]\Vert_1$ of two projected $X$ and $Z$ operators with distance $|j-k|$, such that commutativity is approximately restored at large distances. By construction, the global map exhibits periodic boundary conditions, while the local map does not. Using Haar-random disentanglers/isometries yields qualitatively similar results discussed in Appendix \ref{app:ising}.
The map $V_\text{local}$ implemented by the local MERA picture, however, relates two very different types of theories and their states: Unlike in the global MERA, whose $V_\text{global}$ effectively describes a locality-preserving RG map between critical theories at different length and energy scales, $V_\text{local}$ maps local operators on $H_R$ to both local and highly nonlocal operators on $R$, depending at which time $t$ the operator is inserted on the horizon.
The effect of applying the projector $P=V^\dagger V$ on a critical theory for $V=V_\text{global}$ results in a projection to its low-energy subspace. We can confirm numerically in Appendix \ref{app:ising} that only the first $\sim 2^n$ eigenstates are preserved with meaningful norm. However, choosing a $V=V_\text{local}$ leads to a much slower loss in fidelity as one moves to higher-energy states, thus approximately preserving a large part of the low-energy subspace of the theory (in our example, the critical Ising model). In a quantum gravity interpretation of this model, the effective Hilbert space is thus softly truncated at high energies, as is generally expected due to black hole contributions.

Interestingly, a similar construction using non-isometric maps in Refs.\ \cite{akers2021quantum,akers2022black} produced a similar truncation for the Hilbert space of the black hole interior, but with regard to states of high \emph{complexity} rather than energy.
Using Haar-random projections, it was found that the complexity needed for operations to reach the null space (or kernel) is exponential in the entropy of the black hole. This permits the effective theory to remain approximately valid for all subexponential processes. Similarly, we can easily identify processes that interpolate between physical states and null states in MERA. This can be informative as it tells us how complex an operation would be needed before our effective theory predictions break down. In both the global and local models, the null states describe excitations where the isometry is written as a unitary with partial projection onto an ancilla state orthogonal to $|0\rangle$. 
To produce a null state in the MERA circuit, one thus needs to ``undo'' some gates in the future lightcone of any isometry, change its ancillary projection, and apply the gates in reverse.
Thus if the complexity of a single tensor is $C$, then the complexity to reach any null state is $\leq kC$ where $k$ is upper bounded by the spacetime volume (in Hubble units) of the future lightcone emanating from the altered ancilla and terminating at the late time cutoff. The complexity $C\sim \exp(S_{dS})$ if the tensors themselves are Haar-random, but can be $\mathrm{poly}(S_{dS})$ if they are generated by some local theory or ordinary time evolution in the EFT. 

\vspace*{-3mm}
\section{Discussion}\label{sec:discussion}
Our work proposes a relationship between the Hilbert space dimension verification problem in quantum information and the holographic principle in quantum gravity. With a novel construction of overlapping qubits based on a local tensor network, we create approximate qubits that have distance-dependent overlaps as opposed to ones that are pair-wise constant \cite{chao2017overlapping,akers2021quantum}, a relevant property for physical settings. The specific application of non-isometric maps to dS/MERA relates non-local dynamical processes over the fundamental degrees of freedom to quasi-local processes over a larger set of apparent degrees of freedom, reminiscent of Ref.\ \cite{Susskind:2021esx}. This provides an explicit mapping that reconstructs exterior geometry from horizon degrees of freedom in the local MERA model, delivers new insight on de Sitter quantum gravity, and connects with discussions in eternal inflation and cosmology \cite{Dubovsky:2008rf}. The non-isometricity of the Hilbert space map $V$ also indicates that the theory is unitarity-violating on at least one side of the duality, albeit weakly in the well-spoofed limit.
Though not a required feature for producing overlapping qubits, our specific examples of $V$ are \emph{co-isometric}, i.e., with $V^\dagger$ forming an isometry. 
While natural for describing global de Sitter evolution \cite{Cotler:2022weg}, it is unclear if this property is always necessary in Hilbert space truncations due to quantum gravity.

Although our explicit construction of overlapping qubits differs from that in Ref.\ \cite{chao2017overlapping}, which can be unmasked with linearly many operations in $N$, it is worth commenting on the complexity of such a protocol in a larger gravitational context. For instance, a naive protocol in the actual universe that involves volumetrically many apparent EFT degrees of freedom has the potential to create significant gravitational back-reaction by forming black holes. Therefore it is entirely reasonable from gravitational considerations that operations polynomial in $N$ can be used to verify the dimension of the fundamental Hilbert space. However, the precise scaling is likely relevant for a careful consistency check, which would constitute an interesting future direction.

The idea of spoofing opens up new directions of reformulating quantum field theories in the overlapping qubit language by relaxing the commutation relations for spacelike separated operators in canonical quantization and exploring their phenomenological implications in particle physics and cosmology. When applied to quantum gravity, it remains open whether the overlaps introduced by gravitational dressing can resolve the degree of freedom counting problem in holography beyond AdS/CFT. 

From a complexity theoretic point of view, actual experiments and physical processes are problems that can be solved or verified with polynomial (time or space) complexity (e.g. NP, QMA). However, for problems not belonging to such classes, many distinct processes may appear indistinguishable when a system is probed with limited computational resources\cite{JiLiuSong,Bouland:2022ovo,akers2022black}. The overlapping qubit approach adds to the list yet another instance, where the resource-intensive task is the verification of the exact commutativity between space-like separated operators. 
It thus contributes to an emerging program of understanding quantum gravity from an accuracy- and resource-limited perspective.

\prlsection{Acknowledgments} We  thank Scott Aaronson, Chris Akers, Vijay Balasubramanian, Adam Brown, Oliver Friedrich, Elliott Gesteau, Philipp H\"ohn, John Preskill, Suvrat Raju, Leonard Susskind, Brian Swingle, Yu Tong, Thomas Vidick, Jinzhao Wang, and Zhenbin Yang for interesting comments and discussions.

AJ was supported by the Simons Collaboration on It from Qubit, the US Department of Energy (DE-SC0018407), and the Einstein Research Unit ``Perspectives of a quantum digital transformation''. C.C. acknowledges the support by the U.S. Department of Defense and NIST through the Hartree Postdoctoral Fellowship at QuICS, the Air Force Office of Scientific Research (FA9550-19-1-0360), and the National Science Foundation (PHY-1733907).

\appendix

\section{Remark 1 and Related Discussions}\label{app:proof}
Let us start with
$$\prod_i^{m\leq M}Q^{(i)}_p|\psi_p\rangle = V Q^{(m)}V^{\dagger}\dots V Q^{(1)}V^{\dagger}V|\psi\rangle.$$ Since $PS|\psi\rangle \approx S|\psi\rangle~, \forall S$, the above expression trivially simplifies by applying this condition repeatedly and we have $\prod_i^{m\leq M}Q^{(i)}_p|\psi_p\rangle \approx V\prod^{m\leq M}_i Q^{(i)}|\psi\rangle=V \mathcal{Q}|\psi\rangle.$

Since $\langle \psi_p|\prod_i^{m\leq M}Q^{(i)}_p|\psi_p\rangle = \langle \psi|P \mathcal{Q}|\psi\rangle$ and that $\langle \psi|P\approx \langle \psi|$, it follows that 
$$\langle \psi_p|\prod_i^{\leq M} Q_p^{(i)}|\psi_p\rangle\approx \langle \psi|\prod_i^{\leq M} Q^{(i)}|\psi\rangle.$$

These conditions are sufficient for identifying well-spoofed processes, but are not necessary. 

For example, in the construction of Akers and Pennington \cite{akers2021quantum}, the inner product between vectors $|\psi\rangle,|\phi\rangle$ with support 
 mostly in $ker(P)$ is considered (instead of vectors  satisfying $P|\psi\rangle \approx |\psi\rangle$  as in our construction).
 
Then we see in their construction, that thanks to the random nature of $V$, with high probability
\begin{equation}
    \langle \phi_p|\psi_p\rangle = \sum_i \lambda_i \langle\phi|i\rangle\langle i|\psi\rangle \approx \langle\phi|\psi\rangle,
\end{equation}
where $|i\rangle$ are the eigenstates of $P$ with eigenvalue $\lambda_i$. 

Here we do not provide a rigorous account for the necessary condition but seek to build up some intuition. For any set of operators $\{Q^{(i)}, i=1,\dots M\}$, 

\begin{align*}
    &\langle \psi_p|Q^{(M)}_p\dots Q^{(1)}_p|\psi_p\rangle = \langle \psi|P Q^{(M)} P\dots P Q^{(1)} P|\psi\rangle \\
    =& \sum_{i_j;j=0,\dots M} \prod_{j=0}^M \lambda_{i_j}\langle \psi|i_{M}\rangle\langle i_M| Q^{(M)}|i_{M-1}\rangle\dots\\
    &\times\langle i_1|Q^{(1)}|i_0\rangle\langle i_0|\psi\rangle
\end{align*}
One can think of this expression as a sum of paths with complex weights from $j=0$ to $j=M$ where at each ``time'' $j$ the trajectory can take on different values by inserting $P$ instead of $I$ in a conventional path integral. As a result, some path will terminate when they reach the kernel while others stay in the orthogonal complement. For instance, if we take identical operators $Q\sim \exp(i\Delta t H)$ and $\lambda_i\in\{0,1\}$, this is nothing but the usual propagator but without summing over paths that enter the kernel.

For fixed $V$ and $|\psi\rangle$, the search of $Q^{(i)}$s amounts to solving an approximate constraint satisfaction problem where each $\langle \psi_p|Q^{(M)}_p\dots Q^{(1)}_p|\psi_p\rangle \approx \langle \psi|Q^{(M)}\dots Q^{(1)}|\psi\rangle$ provides a constraint on the matrix elements of $Q^{(i)}$. For a simple example, consider a one-point function $\langle \psi_p|Q_p|\psi_p\rangle\approx \langle \psi|Q|\psi\rangle$ for fixed $V, |\psi\rangle$ and some operator $Q$ with $\lambda_i=1$ for $i=1,\dots,\Lambda$ and $0$ otherwise.  
\begin{equation}
    \sum_{i,k} \langle \psi|i\rangle\langle i|Q|k\rangle\langle k|\psi\rangle = \sum_{i,k=1}^{\Lambda} \psi_i Q_{ik} \psi_k \approx \langle \psi|Q|\psi\rangle.
\end{equation}
This is an approximate constraint on the matrix elements of possible operators $Q$ for which the expectation value is well-spoofed.

\section{Numerics on Spectral Properties}\label{app:spec}

\begin{figure*}
    \centering
    \includegraphics[width=0.8\linewidth]{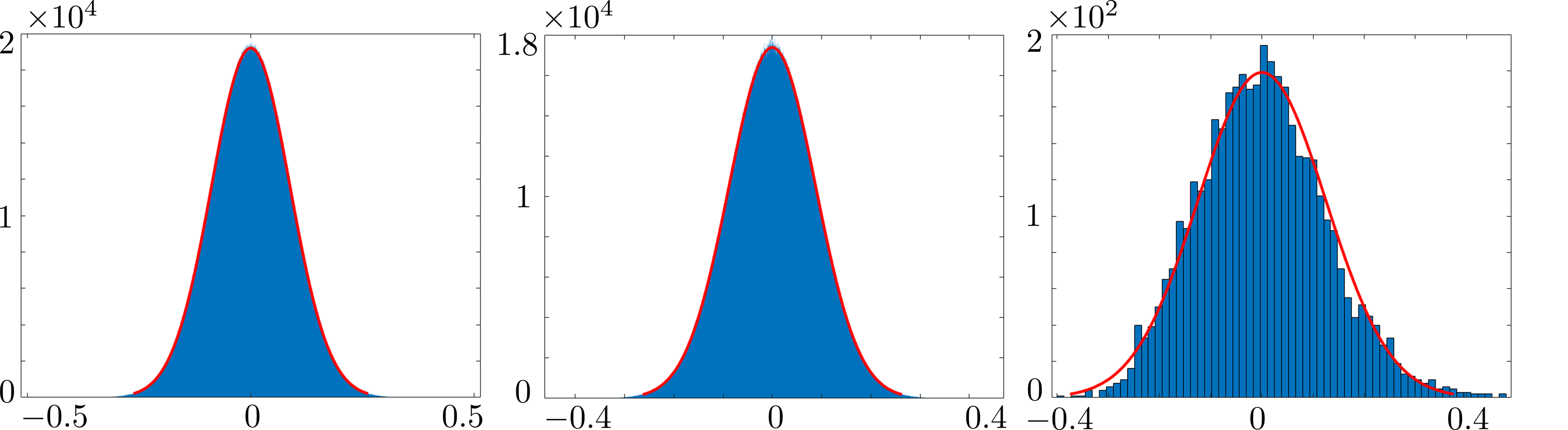}
    \caption{Plotting the distribution of off-diagonal elements in $T_{ij}$ for $N=12, n=6$. The raw (unfitted) variances of the distributions are $\sigma^2_{Re;off}\approx 0.077, \sigma^2_{Im;off}\approx 0.077, \sigma^2_{diag}\approx 0.0156$. The raw means are $-1.6892e-05, 3.2195e-20, 9.8229e-17$ respectively for the off diagonal real, imaginary parts and the diagonal. Red curves are Gaussian fits. The fitted parameters are $(-1.68924e-05,0.0877034),(-3.21951e-20,0.0876864),(9.82287e-17,0.124988 )$ respectively for these 3 data sets.}
    \label{fig:GUE}
\end{figure*}

Here we consider spectral properties of $T_{ij}$ and related operators used for commutators. 
Recall that $T_{ij}=(V^{\dagger}V)-I$. For $V$s that are proportional to Haar random projections like \cite{akers2021quantum}, the individual matrix elements are Gaussian distributed (Figure~\ref{fig:GUE}) which we confirm using small size numerics. Note that however, different matrix elements are correlated and the ensemble generated by different $V$s is not a Gaussian unitary ensemble. This can be seen by noting that $\spec(T)$ has a eigenvalues $2^{N-n}-1$ and $-1$, as required by the singular value decomposition of $V$, i.e., $\spec(P)=\spec(V^{\dagger}V)$ has eigenvalues $2^{N-n}$ for $2^n$ eigenvectors and $0$ for the rest.

\begin{figure*}
    \centering
    \includegraphics[width=0.8\linewidth]{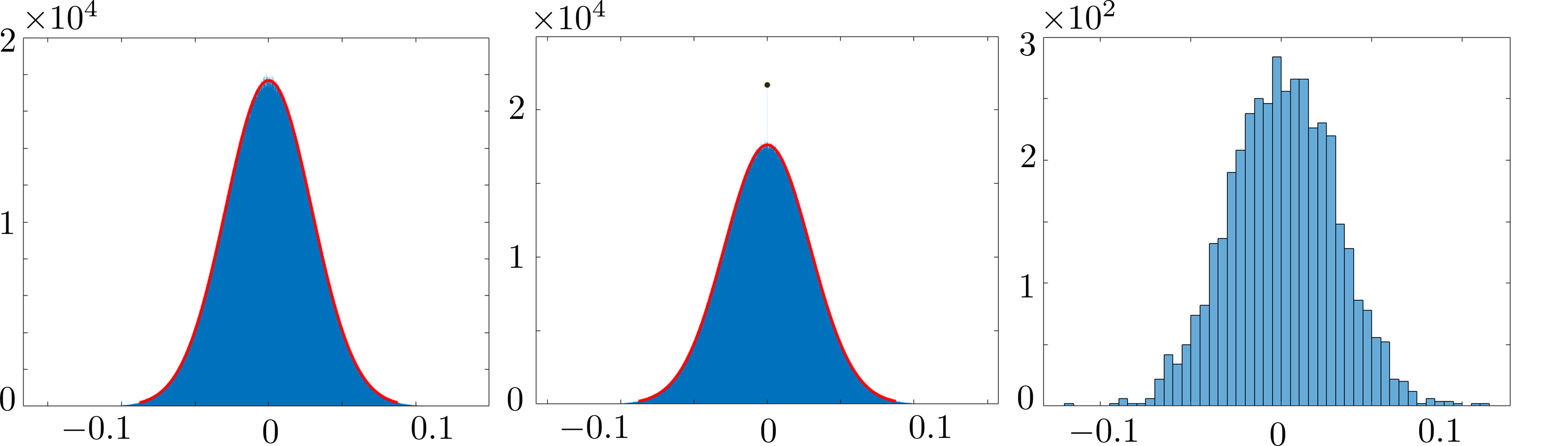}
    \caption{From left to right: distribution of $X_1TZ_2'$ off diagonal real components, off diagonal imaginary components, and diagonal real components. Imaginary diagonal components have similar distribution. Note the peak in the middle figure near 0 which is barely visible due to its bin width but has height $21696$. }
    \label{fig:2pthaar}
\end{figure*}

The behaviour of the commutator, however, is somewhat different. Let us first examine the behaviour of two point functions. Recall that the non-trivial contribution to the projected commutation relation is proportional to 
$VQTQ'V^{\dagger}$ where we assume $Q,Q'$ to be local operators acting on different qubits in $\mathcal{H}_N$. By choosing $Q=X_1, Q'=Z_2$ in our numerics, we note the distributions for $QTQ'$ appear similar, except the off-diagonal imaginary parts have a delta function-like peak near 0. Also the diagonal elements now are complex. The delta-function-like peak is likely related to the local Paulis shifting the diagonal of $T$, which is real. This introduces $2^n$ identical $0$s in the distribution of imaginary components whereas they are non-zero for the real components. Hence the peak only appears in one of the distributions (Figure~\ref{fig:2pthaar}).

\begin{figure*}
    \centering
    \includegraphics[width=0.8\textwidth]{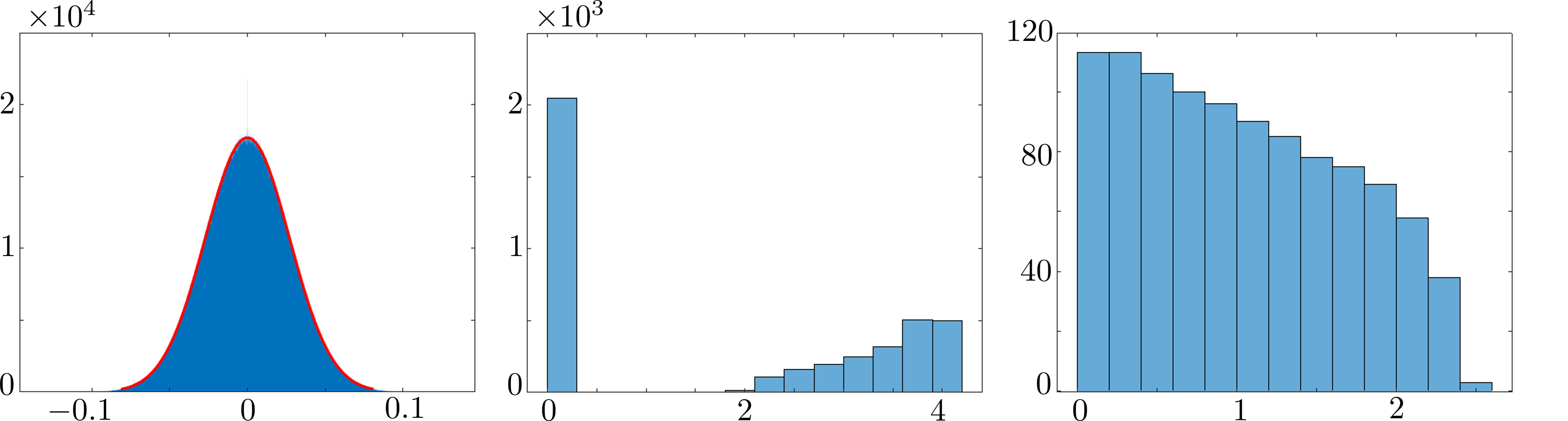}
    \caption{From left to right: imaginary off-diagonal components (note the spike at 0), $|eigenvalue|$ distribution of $C$, $|eigenvalue|$ distribution of $VCV^{\dagger}$. The other components of $C$ are have similar distribution to the 2 point function. The real eigenvalue distributions of $iC$ and $iVCV^{\dagger}$ are symmetric with respect to the y axis, hence the magnitude plot is sufficient. }
    \label{fig:comm_haar}
\end{figure*}

For the commutator, the non-trivial contribution comes from $C=QTQ'-Q'TQ$ except its diagonal is purely imaginary now. This is also understandable because we have chosen $Q,Q'$ to be Hermitian. Therefore, $iC$ is Hermitian. The probability distributions for its diagonal and off-diagonal elements are similar to those of the 2 point function, i.e. , they are mostly Gaussian except for the off-diagonal imaginary elements which have a sharp delta function peak near 0. The eigenvalue distributions for $VCV^{\dagger}$ is more interesting. It is not clear if it is still supposed to follow the semi-circle law. Nevertheless, we do observe a distribution that is mostly concentrated around 0 (Figure~\ref{fig:comm_haar}).

Also for comparison, we can consider a map $V$ that is not Haar random, but is given by the rescaled projection onto the low energy subspace of a 1d critical Ising model
$$H=-\cos\theta \sum_i Z_iZ_{i+1}-\sin\theta \sum_i X_i,$$ where
$V=\Pi_{E<\Lambda}\sqrt{2^{N-n}}$ where $\Lambda$ is the $2^n+1$th energy eigenvalue and the Hamiltonian has periodic boundary condition.

\begin{figure}
    \centering
    \includegraphics[width=0.49\textwidth]{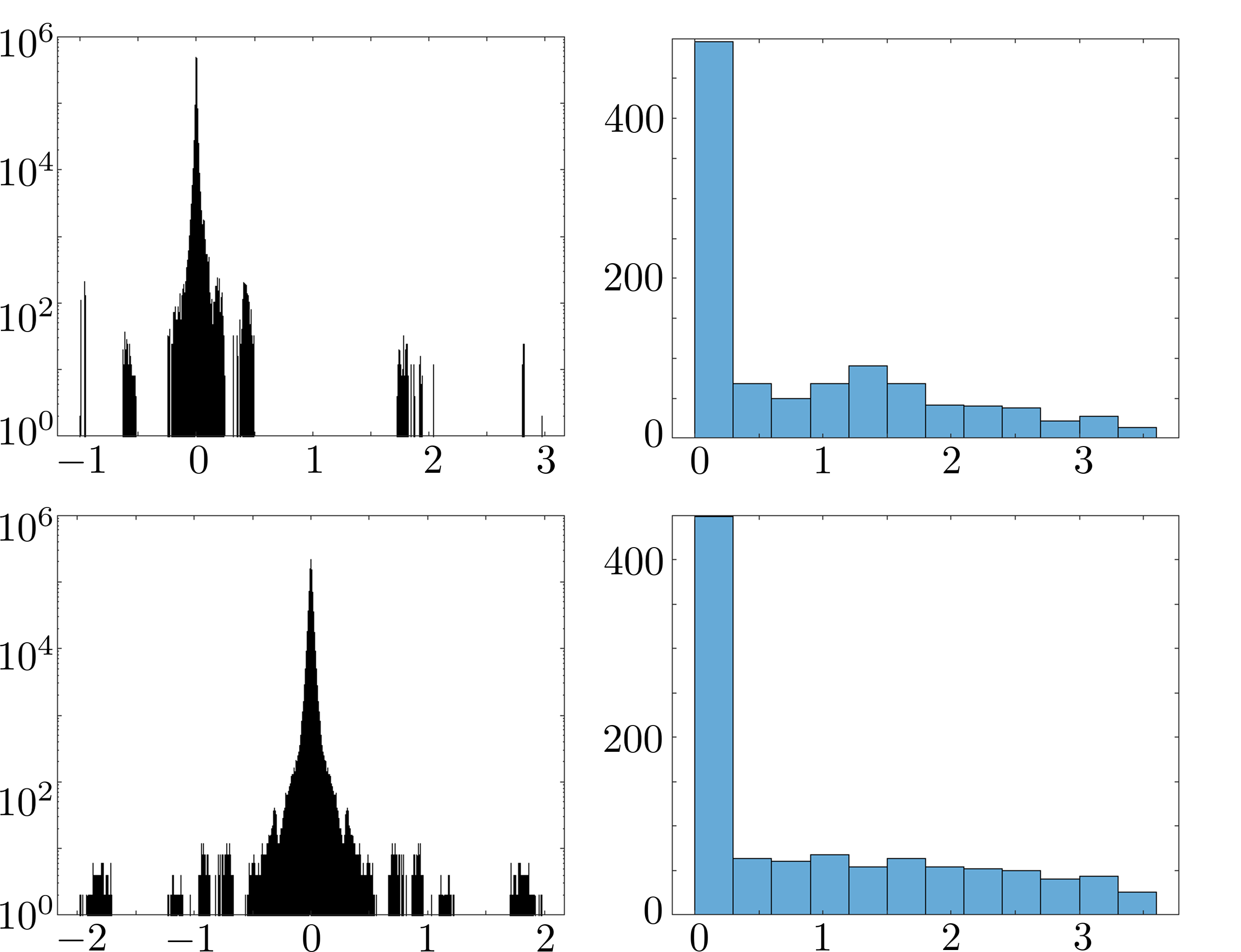}
    \caption{$N=12,n=10$. Left to right: distribution of the value of $C$ when $V$ is given by the rescaled projection onto the low energy subspace; eigenvalue distribution of $|VCV^{\dagger}|$. Top line: non-critical Ising at $\theta=\pi/5$. Bottom line: critical Ising at $\theta=\pi/4$. }
    \label{fig:IsingC}
\end{figure}

It is clear then that the distribution of matrix elements is no longer Gaussian (Figure~\ref{fig:IsingC}). Only real components of $C$ are plotted as imaginary components are 0. Nevertheless, the majority of the values are concentrated around $0$ for $C$ and most of the eigenvalues are also vanishing, hence indicating that the local physics is approximately preserved for most states in the (physical) Hilbert space.
 
Interestingly, if we examine the eigenstates of $|VCV^{\dagger}|$, the states with smaller eigenvalues (i.e. better approximations of the vanishing commutation relations) also tend to have lower energies compared to the ones that have larger violations. The energy here is computed with respect to the projected Ising Hamiltonian $VH_{\rm Ising}V^{\dagger}$. As the projected Hamiltonian need not generate the dynamics on the fundamental Hilbert space, we will refer to this as pseudo-energy instead. More precisely, let $|\phi_{k}\rangle=|\lambda^{|i-j|,k}_p\rangle$ be the $k$-th eigenvector of the commutator $[X_p^{(i)},Z_p^{(j)}]$  with eigenvalue $\lambda_{|i-j|}(k)$ where we have arranged the eigenvalues in ascending order. Then define $$\omega_{|i-j|}(k)= \langle \lambda^{|i-j|,k}_p|VH_{\rm Ising}V^{\dagger}|\lambda^{|i-j|,k}_p\rangle$$ as the pseudo-energy. 
Then the relation between the eigenvalue $\lambda_{\Delta}(k)$ and $ \omega_{\Delta}(k)$ is shown in Fig.~\ref{fig:ising-dep}b where we have also averaged overall positions $i$ with fixed $\Delta=|i-j|$. For this plot, $N=12, n=6$ and $\Delta=5$. Most states preserve the vanishing commutation, but large violation occurs above an energy threshold.

This is intuitive in that we expect lower energy processes to preserve the locality better, while higher energy ones can ``back-react'' more and hence incurring a bigger correction to the commutation relation. Also note the states with a wide range of different energies preserve the commutation relation.

\begin{figure}[t]
    \centering
    \includegraphics[width=0.9\linewidth]{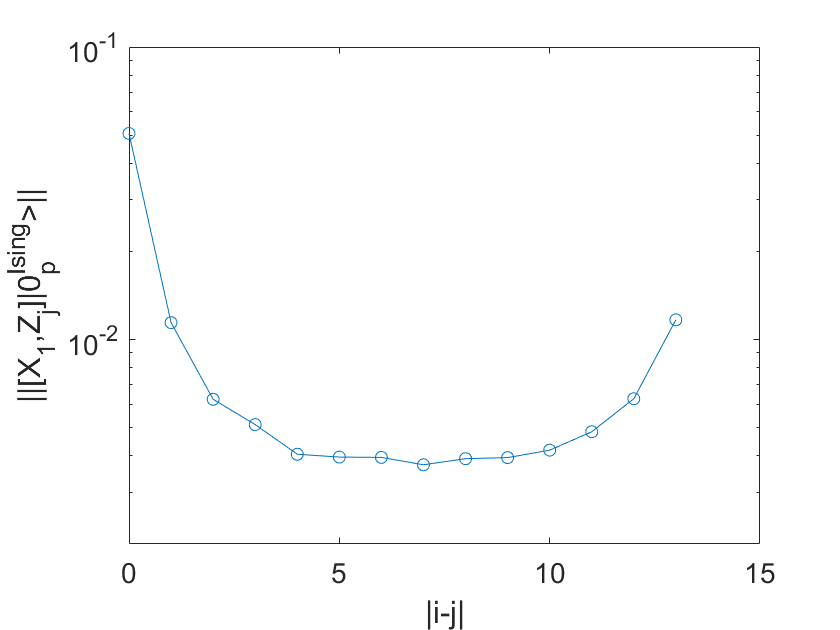}
    \caption{Value of $\Vert[X^{(i)}_p,Z_p^{(j)}]|0^{Ising}_p \rangle\Vert/\Vert X^{(i)}_p\Vert \Vert Z_p^{(j)}\Vert$ plotted with respect to $|i-j|$ for $i=1$. $|0^{Ising}_p\rangle$ is the projected ground state of a 1d Ising model over $N=14$ qubits with periodic boundary condition to a smaller $n=6$ qubit Hilbert space using global energy truncation. The on-site non-commutation is approximately 12 times the smallest overlap.}
    \label{fig:gscomm_ising}
\end{figure}

As a reference, the state-dependent commutator norm $\Vert[X_p^{(i)},Z^{(j)}_p]|\psi_p\rangle\Vert$
also has distance dependence (Fig.~\ref{fig:gscomm_ising}), like the full commutator norm. Unsurprisingly, it has a smaller overlap compared to the full norm (Fig.\ 3 in the main text) as the value is taken with respect to a particular state.

\section{Commutator in global dS MERA}
\label{app:tij}

\begin{figure}[t]
    \centering
    \includegraphics[width=0.48\textwidth]{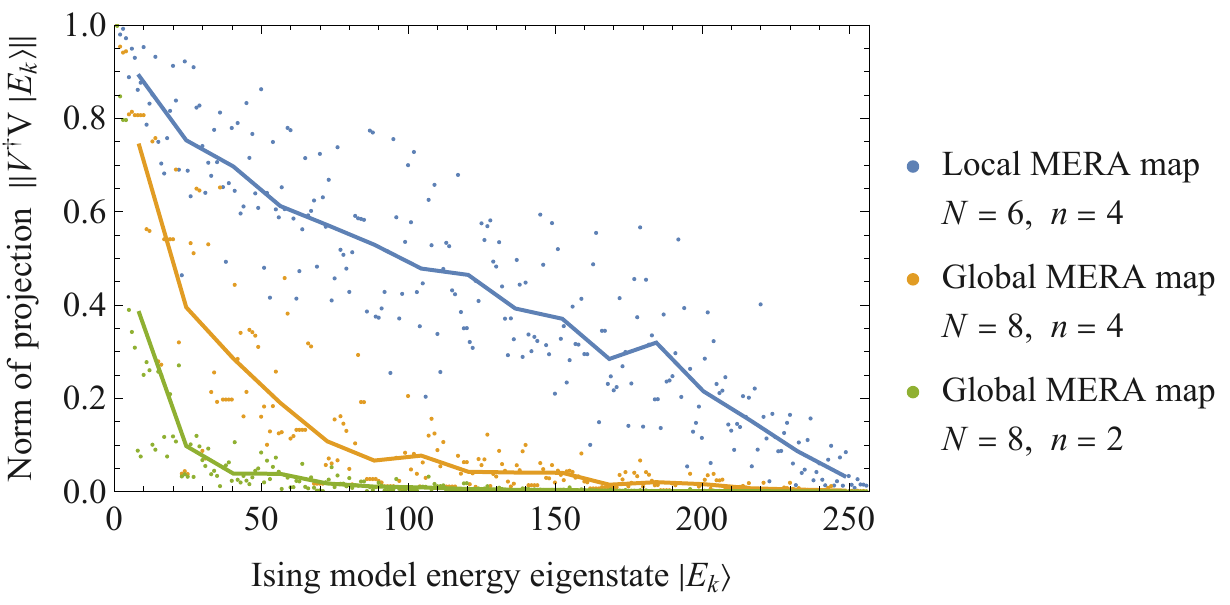}
    \caption{Norm of energy eigenstates of the 8-spin critical Ising model, after projection onto the fundamental Hilbert space of the $N=6,n=4$ local model (times $\id^{\otimes 2}$ acting on two sites) as well as the $N=8,n=4$ and $n=2$ global model. Eigenstates are ordered by their energy eigenvalues. The solid curve shows the average fidelity of blocks of 16 eigenstates. 
    }
    \label{fig:energy_trunc}
\end{figure}

First recall that for two observables $O(i),O(j)$ separated by $|i-j|$ sites on the most UV, $T$-th, ``future infinity'' layer, their ``coarse-grained'' IR versions are given by the superoperators induced by their past causal cone. It is clear that $[O(i)_p,O(j)_p]|\psi_p\rangle=0$ if their past causal cones on $V$ do not intersect.

Other than isolated choice of $i,j$ where the past causal cones never intersect because of the lack of translational symmetry in binary MERA, it generally takes $\sim \log_2 |i-j|$ layers before their past causal cones intersect.  After these causal cones merge, the subsequent coarse-graining will act on these operators as ascending superoperator with an operator spectrum $\{\lambda_k\leq 1\}$ where $\lambda_0=1$ is generically unique for random unitaries. The repeated applications of these superoperators $\mathcal{E}$ slowly shave off the effective target Hilbert space dimension for the remaining $T-\log_2|i-j|$ layers of coarse graining, as the support over smaller eigenvalues will be exponentially suppressed with the number of times the superoperator is applied. 

This action of $\mathcal{E}$ acts like a truncation map, but with a soft cut-off. $L$ applications of $\mathcal{E}$ does the following in the eigenbasis
\begin{equation}
    \mathcal{E}^L\rightarrow 
    \begin{pmatrix}
    \lambda_0^L & 0 & 0 & \dots\\
    0 & \lambda_1^L & 0 & \dots\\
    0 & 0 & \lambda_2^L & \dots\\
    \vdots & \vdots & \ddots & \ddots
    \end{pmatrix}
\end{equation}
with $\lambda_1\geq \lambda_2\geq \dots$. 
Instead of introducing a hard cut off above some $\lambda_k$, this operator suppresses part of the spectrum with each iteration. Because of this exponential suppression, let us approximate $\mathcal{E}^L:N \rightarrow M_L$ as a non-isometric map where $M_L,N$ are the corresponding operator Hilbert spaces and $\dim N \approx \exp(\eta L)\dim M_L$. If we take a generic representative of one such map, it is a random projection that preserves approximate orthogonality. From concentration of measure \cite{chao2017overlapping}, we know that there exists an embedding of $S_{dS}\sim\log N\sim \exp(\epsilon^2_L \log M_L)$ qubits into $\log M_L$ exact qubits with $O(\epsilon_L)$ overlap. For random projection, this only occurs with high probability. At $L=0$, $\log N= 2\log M_0$ in a binary MERA. Since $L=T-\log|i-j|$, there exists a Johnson-Lindenstrauss-type mapping such that 
\begin{align}
    &\Vert[O^{(i)}_p,O^{(j)}_p]\Vert^2<\epsilon^2 \\
    \textrm{where}~&\epsilon^2\sim  \frac{\log S}{(-\eta L+S)} \approx \frac{\log S}{\eta\log|i-j|+S-\eta T}.
\end{align} 

The largest overlap occurs when $L\sim T$, where the two operators started off in the same static patch. Therefore $\epsilon \sim O(1)$ when $T\sim (S-\log S)/\eta$. Note that this is an estimate for the upper bound for the commutator norm, and therefore a lower bound for the cut off of time $T$ after which locality breaks down. 

This time estimate is unsurprising, because there exist known mappings \cite{chao2017overlapping} where $N$ constant pair-wise overlapping qubits can be achieved with $n=O(\frac{1}{\epsilon^2}\log(N))$ non-overlapping ones. While it would limit the time of exponential expansion to $t\sim n\sim S_{dS}$, from a quantum cosmology perspective, the above requirement is still too strong because the Universe is described by a single state vector.
Therefore we are more interested in the overlap with respect to a particular state, which we look at now.

Note that the channel $\mathcal{E}^L$ is produced by a state level non-isometric map supplied by the MERA tensor network such that $W:\mathcal{H}_{L_{int}}\otimes \mathcal{H}_{L_{hor}}\rightarrow \mathcal{H}_{0_A}$ maps from the interior and horizon degrees of freedom of the causal patch at $t=L$ to a subregion $A$ of the $\Sigma_0$ slice after the past causal cones of $O^{(i)},O^{(j)}$ merge. This map does not treat all inputs equally as they are introduced at different layers. Like before, let us suppose that the action of this map on the $\mathcal{H}_{L_{int}}$ subspace can be approximated by a random projection $W':\mathcal{H}_{L_{int}}\rightarrow\mathcal{H}_{M_L}\subset \mathcal{H}_{0_A}$ such that $d_{M_L}=|\mathcal{H}_{M_L}|=2^{-\eta' L}|\mathcal{H}_{L_{int}}|\sim 2^{-\eta'L} e^{S}$. Let $d_{L_{int}}=|\mathcal{H}_{L_{int}}|$. Using a random projection, this allows us to embed $\mathcal{H}_{L_{int}}$ into $\mathcal{H}_{M_L}$ with $d_{L_{int}} \sim \exp(\epsilon_L'^2 d_{M_L})$.
It is easy to show that $\Vert [O(i)_p,O(j)_p]|\psi_p\rangle\Vert  \leq   \epsilon'\Vert T_p^{[O(i),O(j)]}|\psi_p\rangle\Vert $ where the matrix elements of $T^{[O(i),O(j)]}$ is of at most $O(1)$ with high probability and
$$\epsilon'^2 \approx \frac{\log(S)e^{-S}e^{\eta'T}}{|i-j|^{\eta'}}.$$ Thus $\epsilon'$ becomes $O(1)$ for $T\sim (S-\log\log S)/\eta'$.

Therefore, we expect the overall estimate for locality to break down in a time scale similar to the state-independent case. Furthermore, one should expect $\Vert [Q_i,Q_j]|\psi_p\rangle\Vert \lesssim O({\tilde{\epsilon}}/{|i-j|^{\eta'/2}})$ for fixed $S,T$ to satisfy a power law as long as the approximation for the map $W$ is correct (Fig~\ref{fig:gscomm_ising}).

This is not sufficient to show that the norm is small, however, as $T_p$ is generally not unitary. We would also need there to exist $|\psi_p\rangle$ such that $\Vert T_p^{[O(i),O(j)]}|\psi_p\rangle\Vert $ does not grow exponentially with $S$. Although numerically we can verify that $\spec(T_p^{[O(i),O(j)]})$ has a large number of low lying eigenvalues (Appendix~\ref{app:spec}) where the minimum eigenvalue appears stable at different $S$'s, it is insufficient for any asymptotics. There is also no reason to expect that random tensors are physically relevant other than technical convenience. We leave a full analysis of $T_p^{[O(i),O(j)]}$'s spectral properties for local operators $O$ to future work.

Instead of $W'$, we can construct a slightly different map $V:\mathcal{H}_{L_{int}}\rightarrow \mathcal{H}_{M_L}$ with a bit more structure by choosing a different set of disentanglers and isometries.

Let us define
\begin{align*}
    \mathcal{D}_0&=\{|\psi\rangle\}\subset\mathcal{H}_{L_{int}},\\ \mathcal{D}_1&=\operatorname{span}\{Q_i|\psi\rangle, \forall~Q_i\}\cap \mathcal{D}_0^{\perp}, \dots \\ \mathcal{D}_M&=(\bigoplus_{i=0}^{M-1}\mathcal{D}_i)^{\perp}\\
    &\cap \operatorname{span}\{Q_{i_1}\dots Q_{i_M}|\psi\rangle, \forall~Q_{i_j}: i_j\ne i_k~\mathrm{if}~j\ne k\}.
\end{align*} 
It is helpful to think of $\mathcal{D}_i$ as the space of states where the computational basis states have Hamming weight $i$. Consider a non-isometric map $V=\bigoplus_i V_i$ where each $V_i$ is a random projection from $\mathcal{D}_i$ to $\mathcal{D}_i'\subset \mathcal{H}_n$ such that $V^{\dagger}V$ preserves the vector inner product approximately within each block but exactly across different blocks. Define $\mathcal{C}_r=\bigoplus_{i=0}^{r}\mathcal{H}_i$ and $C_r = |\mathcal{C}_r|$. 
Using this map, we embed $\mathcal{C}_{M}$ into a $O(e^{S-\eta'T})$ dimensional subspace of the physical Hilbert space $\mathcal{H}_{0_A}$ such that orthogonality within each subspace $\mathcal{H}_i$ is approximately preserved using the JL theorem up to overlap bounded by $\epsilon$, and have the remaining states mapped to the null space. By the direct sum, orthogonality across different subspaces are exactly preserved.

Then 
\begin{align*}
    &Q_p^{i_M}\dots Q_p^{i_2}Q_p^{i_1}|\psi_p\rangle = Q_p^{i_M}\dots VQ^{i_2}V^{\dagger}V Q^{i_1}|\psi\rangle\\
    &=Q_p^{i_M}\dots VQ^{i_2} Q^{i_1}|\psi\rangle + \epsilon Q_p^{i_M}\dots VQ^{i_2} T Q^{i_1}|\psi\rangle\\
    &=V (Q^{i_M}\dots Q^{i_1}|\psi\rangle + \sum_{k=1}^{M-1}\epsilon^k O_k[Q,T]|\psi\rangle)
\end{align*}
where $O_k[Q,T]$ is the set of all operator strings consisting of $Q^{i_M}\dots Q^{i_1}$ where total $k$ $T$ operators are inserted into the $M-1$ gaps between $Q$ operators. Only a single $T$ may be inserted for each gap. 
Note that $$\Vert O_k[Q,T]|\psi\rangle\Vert \leq C_{M-1}^k$$ because $T$ by construction is block diagonal, hence $T$ applied to any state in $\mathcal{C}_r$ never maps it to a state outside this subspace, unlike the fully Haar random construction. 

Since $|\mathcal{D}_i|={S\choose i} \sim S^i$, $C_{M-1}\leq M S^{M-1}$ which is polynomial for bounded $M$. Then $$\epsilon C_{M-1} \lesssim \frac{S^{M-1} M^{3/2}\sqrt{\log S}}{e^{(S-\eta'L)/2}}\ll 1,$$ for fixed $M$, $S\gg 1$ and $L\lesssim (S-O(M\log S)-\log M)/\eta'$. Recall that $V^{\dagger}V|\psi\rangle = |\psi\rangle$, therefore any $M$-point function can be approximated with the above restrictions
\begin{align*}
    &|\langle\psi_p|Q^{i_M}_p\dots Q^{i_1}_p|\psi_p\rangle- \langle \psi|Q^{i_M}\dots Q^{i_1}|\psi\rangle| \\
    \leq& \sum_{k=1}^{M-1}|\epsilon^k \langle\psi|V^{\dagger}V O_k(Q,T)|\psi\rangle|
    \leq\sum_{k=1}^{M-1}\Big(\epsilon{{M-1}\choose k} C_{M-1}\Big)^k \\
    <&\sum_{k=1}^{M-1}(\epsilon M C_{M-1})^k\ll 1.
\end{align*} 
Thus for operators within the same static patch, this approximation is valid up to $T=L\lesssim O(S - M\log S-\log M)$. Because the state-dependent commutator norm $\Vert [Q^{i_1}_p,Q^{i_2}_p]|\psi_p\rangle\Vert $ can be expanded as a sum of 4 point functions, its values are also preserved for the same amount of time before the correction becomes order 1.

For commutators of operators $Q^{i_1}, Q^{i_2}$ that live in different static patches, we have $T=L+\log|i-j|$, and the correction terms are again $O(\epsilon M C_1) \approx \tilde{\epsilon}/|i-j|^{\eta'/2}$ where $\tilde{\epsilon}\approx S\log S \exp(-S/2+\eta'T/2)$. Again, this value becomes order one in time at most linear in $S$.

Assumptions used to produce the above estimates for MERA may fail. For reference, let us also examine a simpler case where we allow constant pairwise overlap using the same type of random map $V:\mathcal{H}_{N_t}\rightarrow \mathcal{H}_n$ globally. This simplification removes the local structure of the MERA and is more similar to the one by \cite{akers2021quantum}. We use it as an estimate to bound the longest possible time that a spoofing of local physics can last if $N$ increases exponentially in time. At time $T$, the apparent Hilbert space consists of $\sim e^T S$ qubits. If we repeat the same exercise and embed the $C_M$ subspace into the physical Hilbert space of dimension $e^S$ using the direct sum of JL mappings, then one can show that 
$$\epsilon \sim e^{-S/2}\sqrt{T+\log S}, ~C_{M}\leq (M+1) e^{MT} S^M.$$ For any $M>1$, the size of the correction scales polynomially with $\epsilon C_{M-1}$, meaning that it is small for at most $T\lesssim S/2M$. This is consistent with our previous estimate for the amount of time needed before such local physics breaks down. However, for $M=1$, $C_{0}=1$ and the exponential multiplicative factor does not enter. Hence a two point function with respect to the special ``vacuum'' state $|\psi\rangle$ has
\begin{align*}
    &\langle \psi_p|Q^{i_1}_pQ^{i_2}_p|\psi_p\rangle = \langle \psi|Q^{i_1}V^{\dagger}VQ^{i_2}|\psi\rangle \\
    = &\langle \psi|Q^{i_1}Q^{i_2}|\psi\rangle +O(\epsilon).
\end{align*}
Hence the correction here only becomes order 1 when $T\sim e^S-\log S$, which roughly coincides with the time it takes for the proposed dS dynamics to reach maximum complexity \cite{susskind2021sitter}.

\section{Critical Ising and Haar Random MERA}\label{app:ising}

In the main text, we consider a MERA tensor network of bond dimension $\chi=2$ with tensors chosen to approximate the critical Ising model. Here we present the details of this construction: As shown in Fig.\ \ref{fig:ds_wedges}, the two types of MERA tensors are 3-leg isometries and 4-leg unitaries (or \emph{disentanglers}). We can represent these tensors by a $2\times4$ matrix $M_i$ and a $4\times4$ matrix $M_d$, respectively. 
The Ising model at its critical point is given by Hamiltonian 
\begin{equation}
\label{EQ_ISING_H}
    H_\text{I} = - \sum_{k=1}^N \left( X_k X_{k+1} + Z_k \right) \ ,
\end{equation}
where $X_k, Z_k$ are Pauli operators acting on the $k$th site in a chain of $N$ spins. Its ground state can be solved using free-fermion  techniques, and so it is convenient to use MERA tensors that are themselves free fermion operators. An ansatz for $M_i$ and $M_d$ that fulfills the isometry/unitary constraint $M_i^\dagger M_i = M_d^\dagger M_d = \id$ and is free-fermionic is given by \cite{Evenbly:2016cly}
\begin{align}
    M_i &= 
    \begin{pmatrix}
        \cos\alpha & 0 & 0 & \sin\alpha \\
        0 & \frac{1}{\sqrt{2}} & - \frac{1}{\sqrt{2}}& 0
    \end{pmatrix} \ , \\
    M_d &= 
    \begin{pmatrix}
         \cos \beta & 0 & 0 & \sin\beta \\
         0 & 1 & 0 & 0 \\
         0 & 0 & -1 & 0 \\
         -\sin \beta & 0 & 0 & \cos\beta
    \end{pmatrix} \ ,
\end{align}
where $\alpha,\beta$ are real angles. Note that our $M_d$ differs slightly from the one proposed in \cite{Evenbly:2016cly}, though both are valid solutions. We also need to specify an initial (central) 4-leg tensor $T^0$. A $T^0$ that represents a free fermion state with minimal energy with respect to \eqref{EQ_ISING_H} can be found analytically \cite{Jahn:2017tls} and has nonzero entries
\begin{align}
    T_{1,1,1,1} &= \frac{1}{N} \ , \\
    T_{2,2,1,1} &= T_{1,2,2,1} = T_{1,1,2,2} = T_{2,1,1,2} = \frac{a_0}{N} \ , \\
    T_{2,1,2,1} &= T_{1,2,1,2} = \frac{b_0}{N} \ , \\
    T_{2,2,2,2} &= \frac{2a_0^2 - b_0^2}{N} \,
\end{align}
where
\begin{align}
    N = \sqrt{1 + 4a_0^2 + 2b_0^2 + (2a_0^2 - b_0^2)^2} \ ,
\end{align}
with $a_0 \approx 0.3066 \ , b_0 \approx 0.2346$. 
Given this initial tensor, one can then find the $\alpha,\beta$ that best approximate the critical Ising ground state. For a MERA with two layers describing $N=16$ sites, we perform a numerical minimization using a matchgate tensor network \cite{Jahn:2017tls} to find $\alpha \approx 1.831, \beta \approx  1.682$. This corresponds to a ground state energy of $\approx - 0.6358$ per site, very close to the exact value $-\frac{2}{\pi} \approx - 0.6366$ for the critical Ising model in the continuum limit.

For both the global and local MERA maps $V:\mathcal{H}_N\to\mathcal{H}_n$ considered in the main text, we can build a projector $P=V^\dagger V$ onto the ``fundamental subspace'' of $\mathcal{H}_N$ using the explicit tensor network form of $V$. We then compute the overlap of this subspace with the energy eigenbasis of the Ising Hamiltonian \eqref{EQ_ISING_H}. The result, shown in Fig.\ \ref{fig:energy_trunc}, is that both the global and local MERA preserve the low-energy part of the spectrum. While the global MERA truncation is sharp (as expected from the IR/UV interpretation of the MERA layers), the local MERA truncation gradually loses fidelity as one moves from lower- to higher-energy states.

\begin{figure}[t]
    \centering
    \includegraphics[width=0.48\textwidth]{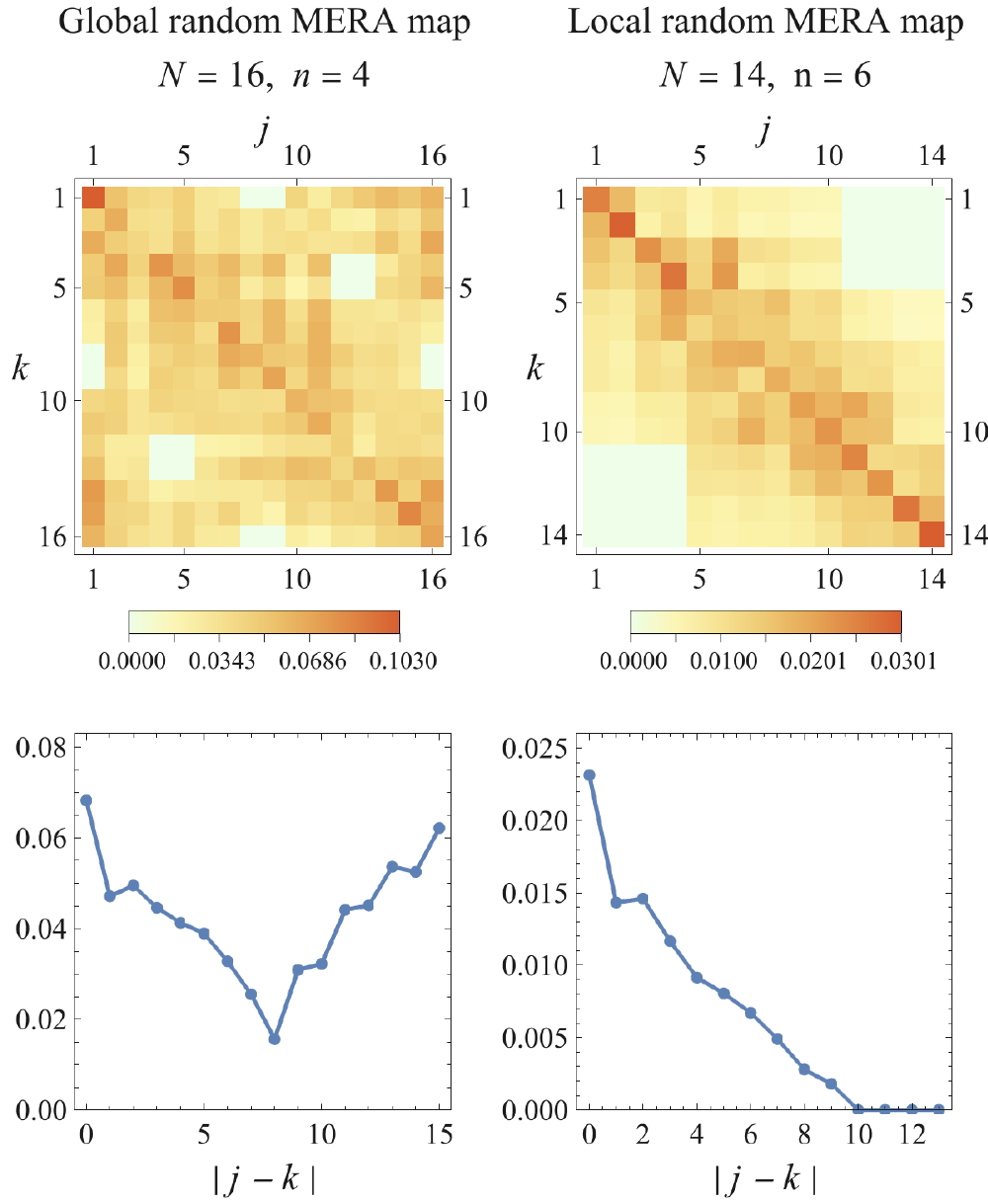}
    \caption{Commutator trace norms $\Vert[\tilde{X}_j,\tilde{Z}_k]\Vert_1 / \Vert \tilde{X}_j\Vert_1 \Vert \tilde{Z}_k\Vert_1$ of projected Pauli operators as in Fig.\ \ref{fig:xz_comm2}, but for the global and local MERA map with Haar-random isometries and disentanglers. The plots are averages of five samples, with tensors chosen locally random in each sample. The second row shows the average decay with distance $|j-k|$. 
    }
    \label{fig:xz_comm2}
\end{figure}

Rather than the non-interacting Ising model, we consider a second numerical setup in which the MERA tensors are locally chosen Haar-randomly. Specifically, that means that each disentangler is chosen as an independently Haar-random unitary map acting on two qubits, while each isometry is chosen by the projection of such a Haar-random unitary on the $\ket{0}$ state of one of the two output qubits, resulting in an isometric map. 
The norm of the resulting commutators is shown in Fig.\ \ref{fig:xz_comm2}, and follows similar qualitative features as the Ising case, suggesting that the MERA geometry and tensor restrictions, by themselves, ensure an approximate locality in the truncated Hilbert space. 
Note that while we only consider local (bond) dimension $\chi=2$, such a construction can easily be generalized to any larger $\chi$, serving as a toy model for higher-dimensional CFT states.

\section{Entanglement in the static patch}
\label{app:static-patch-ee}

\begin{figure}[ht]
    \centering
    \includegraphics[width=0.6\linewidth]{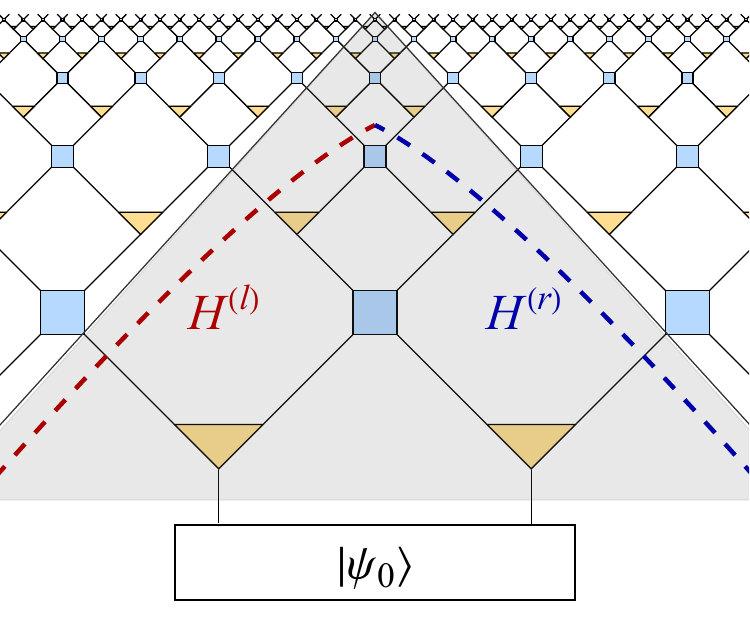} \\[0.2cm]
    \includegraphics[width=0.9\linewidth]{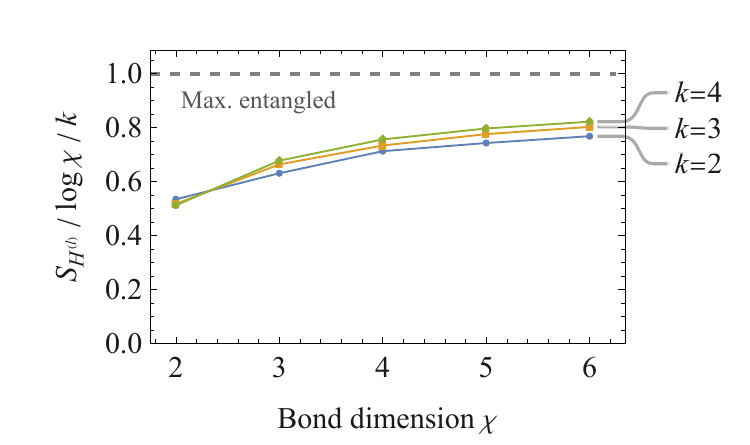}
    \caption{Entanglement along the static patch horizon in the MERA discretization of dS$_2$. We consider an initial state $|\psi_0\rangle=|0\rangle^{\otimes 2}$ and apply two isometries and one disentangler in each layer. We approximate the static patch horizon by terminating after $k$ layers, here shown for $k=2$, and compute the entanglement entropy $S_{H^{(l)}}$ between the left and right half $H^{(l)}$ and $H^{(r)}$. Shown below is the result for Haar-random unitaries and isometries of bond dimension $\chi$, with apparent convergence to maximal entanglement as $\chi,k \to \infty$. 
    }
    \label{fig:static-patch-ee}
\end{figure}

As discussed in the main text, the local MERA map (Fig.2(b) in the main text), when applied to the exterior region between the static patches of two antipodal observers in dS$_2$, relates the Hilbert space on (half of) the static patch horizons to that of a time-slice of the exterior region. This raises the following question: If we choose the MERA tensors to well approximate the ground state of a critical theory, and hence produce a subsystem reduction $\rho_R$ of such a state on a time-slice of the exterior region, what is the dual state $\tilde{\rho}_{H_R} \equiv \tilde{\rho}_R$ on the horizon degrees of freedom produced by applying the local MERA map? This density matrix $\tilde{\rho}_R$ can be constructed by contracting all of the tensors \emph{outside} of the local wedge $\mathcal{W}_R$. By symmetry, this corresponds to contracting the tensors within each static patch. For generic tensors, such as those describing the  bond dimension $\chi=2$ MERA approximation of the critical Ising model, $\tilde{\rho}_R$ will describe a state with complicated entanglement. The entanglement along the horizon can, however, be bounded by $|H_R| \log\chi$, $|H_R|$ being the number of sites along the horizon $H_R$, which we make finite by a time-slice cutoff at finite $t$. This argument is exactly analogous to the Ryu-Takayanagi entanglement entropy bound in previous AdS/MERA proposals \cite{Swingle:2009bg}, where $H_r$ takes the form of the RT surface $\gamma_R$ \cite{Ryu:2006bv}.
As in AdS/MERA, we expect our tensor network to represent gravitational features only in the limit of very large bond dimension $\chi$. Choosing such a limit for random tensors, one finds that the entanglement entropy saturates the upper bound $|\gamma_R| \log\chi$, i.e., exhibits maximal entanglement across minimal cuts $\gamma_R$ \cite{Hayden:2016cfa}.
Applying this observation to our dS picture, we expect that for a MERA tensor network with Haar-random disentanglers and isometries at large bond dimension, the entanglement along $H_R$ will asymptote to $|H_R| \log\chi$. Within each static patch, this implies that the left and right ``half'' of each horizon become maximally entangled with each other, as the minimal cut for each half coincides with the sites themselves. We show the slow convergence to this limit numerically in Fig.\ \ref{fig:static-patch-ee}.
This leads to a horizon state $\tilde{\rho}_R \propto \id$, i.e., the maximally mixed state. This behavior is striking, as it implies that the local MERA map $V$ relates the entanglement spectrum of a CFT ground state (assuming such can be approximated by high bond dimension MERA) to a completely flat one characteristic of maximal entanglement. $V$ thus appears to distill the complicated entanglement between a CFT subregion $R$ and its complement into EPR pairs. This behavior has interesting consequences in the continuum limit, where the partial trace of QFT subregions is ill-defined due to divergences in the entanglement spectrum formally associated with a type III von Neumann algebra. However, it has recently been proposed that gravity effects in de Sitter spacetime effectively reduce the algebra of observables in the static patch to one of type II$_1$, where notions of partial traces and entanglement entropies are well-defined \cite{chandrasekaran2022algebra}. There are some similarities between this approach and our work, in particular our relation between a maximum-entropy state on one side and and a CFT ground state on another, though the latter is located in the exterior region in our model, rather than in the static patch itself. The more precise behavior of the continuum limit of our local and global MERA maps is an interesting area for future work.


\begin{thebibliography}{52}%
\makeatletter
\providecommand \@ifxundefined [1]{%
 \@ifx{#1\undefined}
}%
\providecommand \@ifnum [1]{%
 \ifnum #1\expandafter \@firstoftwo
 \else \expandafter \@secondoftwo
 \fi
}%
\providecommand \@ifx [1]{%
 \ifx #1\expandafter \@firstoftwo
 \else \expandafter \@secondoftwo
 \fi
}%
\providecommand \natexlab [1]{#1}%
\providecommand \enquote  [1]{``#1''}%
\providecommand \bibnamefont  [1]{#1}%
\providecommand \bibfnamefont [1]{#1}%
\providecommand \citenamefont [1]{#1}%
\providecommand \href@noop [0]{\@secondoftwo}%
\providecommand \href [0]{\begingroup \@sanitize@url \@href}%
\providecommand \@href[1]{\@@startlink{#1}\@@href}%
\providecommand \@@href[1]{\endgroup#1\@@endlink}%
\providecommand \@sanitize@url [0]{\catcode `\\12\catcode `\$12\catcode
  `\&12\catcode `\#12\catcode `\^12\catcode `\_12\catcode `\%12\relax}%
\providecommand \@@startlink[1]{}%
\providecommand \@@endlink[0]{}%
\providecommand \url  [0]{\begingroup\@sanitize@url \@url }%
\providecommand \@url [1]{\endgroup\@href {#1}{\urlprefix }}%
\providecommand \urlprefix  [0]{URL }%
\providecommand \Eprint [0]{\href }%
\providecommand \doibase [0]{https://doi.org/}%
\providecommand \selectlanguage [0]{\@gobble}%
\providecommand \bibinfo  [0]{\@secondoftwo}%
\providecommand \bibfield  [0]{\@secondoftwo}%
\providecommand \translation [1]{[#1]}%
\providecommand \BibitemOpen [0]{}%
\providecommand \bibitemStop [0]{}%
\providecommand \bibitemNoStop [0]{.\EOS\space}%
\providecommand \EOS [0]{\spacefactor3000\relax}%
\providecommand \BibitemShut  [1]{\csname bibitem#1\endcsname}%
\let\auto@bib@innerbib\@empty
\bibitem [{\citenamefont {Chao}\ \emph {et~al.}(2017)\citenamefont {Chao},
  \citenamefont {Reichardt}, \citenamefont {Sutherland},\ and\ \citenamefont
  {Vidick}}]{chao2017overlapping}%
  \BibitemOpen
  \bibfield  {author} {\bibinfo {author} {\bibfnamefont {R.}~\bibnamefont
  {Chao}}, \bibinfo {author} {\bibfnamefont {B.~W.}\ \bibnamefont {Reichardt}},
  \bibinfo {author} {\bibfnamefont {C.}~\bibnamefont {Sutherland}},\ and\
  \bibinfo {author} {\bibfnamefont {T.}~\bibnamefont {Vidick}},\ }\bibinfo
  {organization} {8th Innovations in Theoretical Computer Science Conference
  (ITCS 2017)}\ (\bibinfo  {publisher} {Schloss Dagstuhl – Leibniz-Zentrum
  für Informatik},\ \bibinfo {year} {2017})\ pp.\ \bibinfo {pages}
  {48:1--48:21},\ \Eprint {https://arxiv.org/abs/1701.01062} {arXiv:1701.01062
  [quant-ph]} \BibitemShut {NoStop}%
\bibitem [{\citenamefont {Bekenstein}(1981)}]{Bekenstein1}%
  \BibitemOpen
  \bibfield  {author} {\bibinfo {author} {\bibfnamefont {J.~D.}\ \bibnamefont
  {Bekenstein}},\ }\href {https://doi.org/10.1103/PhysRevD.23.287} {\bibfield
  {journal} {\bibinfo  {journal} {Phys. Rev. D}\ }\textbf {\bibinfo {volume}
  {23}},\ \bibinfo {pages} {287} (\bibinfo {year} {1981})}\BibitemShut
  {NoStop}%
\bibitem [{\citenamefont {Bekenstein}(1994)}]{Bekenstein:1993dz}%
  \BibitemOpen
  \bibfield  {author} {\bibinfo {author} {\bibfnamefont {J.~D.}\ \bibnamefont
  {Bekenstein}},\ }\href {https://doi.org/10.1103/PhysRevD.49.1912} {\bibfield
  {journal} {\bibinfo  {journal} {Phys. Rev. D}\ }\textbf {\bibinfo {volume}
  {49}},\ \bibinfo {pages} {1912} (\bibinfo {year} {1994})},\ \Eprint
  {https://arxiv.org/abs/gr-qc/9307035} {arXiv:gr-qc/9307035} \BibitemShut
  {NoStop}%
\bibitem [{\citenamefont {Bousso}(1999)}]{Bousso:1999xy}%
  \BibitemOpen
  \bibfield  {author} {\bibinfo {author} {\bibfnamefont {R.}~\bibnamefont
  {Bousso}},\ }\href {https://doi.org/10.1088/1126-6708/1999/07/004} {\bibfield
   {journal} {\bibinfo  {journal} {JHEP}\ }\textbf {\bibinfo {volume} {07}},\
  \bibinfo {pages} {004} (\bibinfo {year} {1999})},\ \Eprint
  {https://arxiv.org/abs/hep-th/9905177} {arXiv:hep-th/9905177} \BibitemShut
  {NoStop}%
\bibitem [{\citenamefont {'t~Hooft}(1993)}]{tHooft:1993dmi}%
  \BibitemOpen
  \bibfield  {author} {\bibinfo {author} {\bibfnamefont {G.}~\bibnamefont
  {'t~Hooft}},\ }\bibfield  {booktitle} {\emph {\bibinfo {booktitle}
  {{Conference on Highlights of Particle and Condensed Matter Physics
  (SALAMFEST) Trieste, Italy, March 8-12, 1993}}},\ }\href@noop {} {\bibfield
  {journal} {\bibinfo  {journal} {Conf. Proc.}\ }\textbf {\bibinfo {volume}
  {C930308}},\ \bibinfo {pages} {284} (\bibinfo {year} {1993})},\ \Eprint
  {https://arxiv.org/abs/gr-qc/9310026} {arXiv:gr-qc/9310026 [gr-qc]}
  \BibitemShut {NoStop}%
\bibitem [{\citenamefont {Susskind}(1995)}]{Susskind:1994vu}%
  \BibitemOpen
  \bibfield  {author} {\bibinfo {author} {\bibfnamefont {L.}~\bibnamefont
  {Susskind}},\ }\href {https://doi.org/10.1063/1.531249} {\bibfield  {journal}
  {\bibinfo  {journal} {J. Math. Phys.}\ }\textbf {\bibinfo {volume} {36}},\
  \bibinfo {pages} {6377} (\bibinfo {year} {1995})},\ \Eprint
  {https://arxiv.org/abs/hep-th/9409089} {arXiv:hep-th/9409089 [hep-th]}
  \BibitemShut {NoStop}%
\bibitem [{\citenamefont {Bekenstein}(1973)}]{Bekenstein:1973ur}%
  \BibitemOpen
  \bibfield  {author} {\bibinfo {author} {\bibfnamefont {J.~D.}\ \bibnamefont
  {Bekenstein}},\ }\href {https://doi.org/10.1103/PhysRevD.7.2333} {\bibfield
  {journal} {\bibinfo  {journal} {Phys. Rev. D}\ }\textbf {\bibinfo {volume}
  {7}},\ \bibinfo {pages} {2333} (\bibinfo {year} {1973})}\BibitemShut
  {NoStop}%
\bibitem [{\citenamefont {Hawking}(1975)}]{Hawking:1974sw}%
  \BibitemOpen
  \bibfield  {author} {\bibinfo {author} {\bibfnamefont {S.~W.}\ \bibnamefont
  {Hawking}},\ }\bibfield  {booktitle} {\emph {\bibinfo {booktitle} {{Euclidean
  quantum gravity}}},\ }\href {https://doi.org/10.1007/BF02345020} {\bibfield
  {journal} {\bibinfo  {journal} {Commun. Math. Phys.}\ }\textbf {\bibinfo
  {volume} {43}},\ \bibinfo {pages} {199} (\bibinfo {year} {1975})}\BibitemShut
  {NoStop}%
\bibitem [{\citenamefont {Maldacena}(1998)}]{Maldacena:1997re}%
  \BibitemOpen
  \bibfield  {author} {\bibinfo {author} {\bibfnamefont {J.~M.}\ \bibnamefont
  {Maldacena}},\ }\href {https://doi.org/10.1023/A:1026654312961} {\bibfield
  {journal} {\bibinfo  {journal} {Adv. Theor. Math. Phys.}\ }\textbf {\bibinfo
  {volume} {2}},\ \bibinfo {pages} {231} (\bibinfo {year} {1998})},\ \Eprint
  {https://arxiv.org/abs/hep-th/9711200} {arXiv:hep-th/9711200} \BibitemShut
  {NoStop}%
\bibitem [{\citenamefont {Akers}\ \emph {et~al.}(2022)\citenamefont {Akers},
  \citenamefont {Engelhardt}, \citenamefont {Harlow}, \citenamefont
  {Penington},\ and\ \citenamefont {Vardhan}}]{akers2022black}%
  \BibitemOpen
  \bibfield  {author} {\bibinfo {author} {\bibfnamefont {C.}~\bibnamefont
  {Akers}}, \bibinfo {author} {\bibfnamefont {N.}~\bibnamefont {Engelhardt}},
  \bibinfo {author} {\bibfnamefont {D.}~\bibnamefont {Harlow}}, \bibinfo
  {author} {\bibfnamefont {G.}~\bibnamefont {Penington}},\ and\ \bibinfo
  {author} {\bibfnamefont {S.}~\bibnamefont {Vardhan}},\ }\href@noop {}
  {\bibfield  {journal} {\bibinfo  {journal} {arXiv preprint arXiv:2207.06536}\
  } (\bibinfo {year} {2022})}\BibitemShut {NoStop}%
\bibitem [{\citenamefont {Balasubramanian}\ \emph
  {et~al.}(2022{\natexlab{a}})\citenamefont {Balasubramanian}, \citenamefont
  {Lawrence}, \citenamefont {Magan},\ and\ \citenamefont
  {Sasieta}}]{Balasubramanian:2022gmo}%
  \BibitemOpen
  \bibfield  {author} {\bibinfo {author} {\bibfnamefont {V.}~\bibnamefont
  {Balasubramanian}}, \bibinfo {author} {\bibfnamefont {A.}~\bibnamefont
  {Lawrence}}, \bibinfo {author} {\bibfnamefont {J.~M.}\ \bibnamefont
  {Magan}},\ and\ \bibinfo {author} {\bibfnamefont {M.}~\bibnamefont
  {Sasieta}},\ }\href@noop {} {\  (\bibinfo {year} {2022}{\natexlab{a}})},\
  \Eprint {https://arxiv.org/abs/2212.02447} {arXiv:2212.02447 [hep-th]}
  \BibitemShut {NoStop}%
\bibitem [{\citenamefont {Balasubramanian}\ \emph
  {et~al.}(2022{\natexlab{b}})\citenamefont {Balasubramanian}, \citenamefont
  {Lawrence}, \citenamefont {Magan},\ and\ \citenamefont
  {Sasieta}}]{Balasubramanian:2022lnw}%
  \BibitemOpen
  \bibfield  {author} {\bibinfo {author} {\bibfnamefont {V.}~\bibnamefont
  {Balasubramanian}}, \bibinfo {author} {\bibfnamefont {A.}~\bibnamefont
  {Lawrence}}, \bibinfo {author} {\bibfnamefont {J.~M.}\ \bibnamefont
  {Magan}},\ and\ \bibinfo {author} {\bibfnamefont {M.}~\bibnamefont
  {Sasieta}},\ }\href@noop {} {\  (\bibinfo {year} {2022}{\natexlab{b}})},\
  \Eprint {https://arxiv.org/abs/2212.08623} {arXiv:2212.08623 [hep-th]}
  \BibitemShut {NoStop}%
\bibitem [{\citenamefont {Akers}\ and\ \citenamefont
  {Penington}(2021)}]{akers2021quantum}%
  \BibitemOpen
  \bibfield  {author} {\bibinfo {author} {\bibfnamefont {C.}~\bibnamefont
  {Akers}}\ and\ \bibinfo {author} {\bibfnamefont {G.}~\bibnamefont
  {Penington}},\ }\href@noop {} {\bibfield  {journal} {\bibinfo  {journal}
  {arXiv preprint arXiv:2109.14618}\ } (\bibinfo {year} {2021})}\BibitemShut
  {NoStop}%
\bibitem [{\citenamefont {Marolf}\ and\ \citenamefont
  {Maxfield}(2020)}]{Marolf:2020xie}%
  \BibitemOpen
  \bibfield  {author} {\bibinfo {author} {\bibfnamefont {D.}~\bibnamefont
  {Marolf}}\ and\ \bibinfo {author} {\bibfnamefont {H.}~\bibnamefont
  {Maxfield}},\ }\href {https://doi.org/10.1007/JHEP08(2020)044} {\bibfield
  {journal} {\bibinfo  {journal} {JHEP}\ }\textbf {\bibinfo {volume} {08}},\
  \bibinfo {pages} {044} (\bibinfo {year} {2020})},\ \Eprint
  {https://arxiv.org/abs/2002.08950} {arXiv:2002.08950 [hep-th]} \BibitemShut
  {NoStop}%
\bibitem [{\citenamefont {Marolf}(2015)}]{Marolf_2015}%
  \BibitemOpen
  \bibfield  {author} {\bibinfo {author} {\bibfnamefont {D.}~\bibnamefont
  {Marolf}},\ }\href {https://doi.org/10.1088/0264-9381/32/24/245003}
  {\bibfield  {journal} {\bibinfo  {journal} {Classical and Quantum Gravity}\
  }\textbf {\bibinfo {volume} {32}},\ \bibinfo {pages} {245003} (\bibinfo
  {year} {2015})}\BibitemShut {NoStop}%
\bibitem [{\citenamefont {Papadodimas}\ and\ \citenamefont
  {Raju}(2014)}]{Papadodimas:2013jku}%
  \BibitemOpen
  \bibfield  {author} {\bibinfo {author} {\bibfnamefont {K.}~\bibnamefont
  {Papadodimas}}\ and\ \bibinfo {author} {\bibfnamefont {S.}~\bibnamefont
  {Raju}},\ }\href {https://doi.org/10.1103/PhysRevD.89.086010} {\bibfield
  {journal} {\bibinfo  {journal} {Phys. Rev. D}\ }\textbf {\bibinfo {volume}
  {89}},\ \bibinfo {pages} {086010} (\bibinfo {year} {2014})},\ \Eprint
  {https://arxiv.org/abs/1310.6335} {arXiv:1310.6335 [hep-th]} \BibitemShut
  {NoStop}%
\bibitem [{\citenamefont {Donnelly}(2012)}]{Donnelly:2011hn}%
  \BibitemOpen
  \bibfield  {author} {\bibinfo {author} {\bibfnamefont {W.}~\bibnamefont
  {Donnelly}},\ }\href {https://doi.org/10.1103/PhysRevD.85.085004} {\bibfield
  {journal} {\bibinfo  {journal} {Phys. Rev. D}\ }\textbf {\bibinfo {volume}
  {85}},\ \bibinfo {pages} {085004} (\bibinfo {year} {2012})},\ \Eprint
  {https://arxiv.org/abs/1109.0036} {arXiv:1109.0036 [hep-th]} \BibitemShut
  {NoStop}%
\bibitem [{\citenamefont {Donnelly}\ and\ \citenamefont
  {Giddings}(2016)}]{Donnelly:2016rvo}%
  \BibitemOpen
  \bibfield  {author} {\bibinfo {author} {\bibfnamefont {W.}~\bibnamefont
  {Donnelly}}\ and\ \bibinfo {author} {\bibfnamefont {S.~B.}\ \bibnamefont
  {Giddings}},\ }\href {https://doi.org/10.1103/PhysRevD.94.104038} {\bibfield
  {journal} {\bibinfo  {journal} {Phys. Rev. D}\ }\textbf {\bibinfo {volume}
  {94}},\ \bibinfo {pages} {104038} (\bibinfo {year} {2016})},\ \Eprint
  {https://arxiv.org/abs/1607.01025} {arXiv:1607.01025 [hep-th]} \BibitemShut
  {NoStop}%
\bibitem [{\citenamefont {Bao}\ \emph {et~al.}(2017)\citenamefont {Bao},
  \citenamefont {Cao}, \citenamefont {Carroll},\ and\ \citenamefont
  {Chatwin-Davies}}]{bao2017sitter}%
  \BibitemOpen
  \bibfield  {author} {\bibinfo {author} {\bibfnamefont {N.}~\bibnamefont
  {Bao}}, \bibinfo {author} {\bibfnamefont {C.}~\bibnamefont {Cao}}, \bibinfo
  {author} {\bibfnamefont {S.~M.}\ \bibnamefont {Carroll}},\ and\ \bibinfo
  {author} {\bibfnamefont {A.}~\bibnamefont {Chatwin-Davies}},\ }\href@noop {}
  {\bibfield  {journal} {\bibinfo  {journal} {Physical Review D}\ }\textbf
  {\bibinfo {volume} {96}},\ \bibinfo {pages} {123536} (\bibinfo {year}
  {2017})}\BibitemShut {NoStop}%
\bibitem [{\citenamefont {Susskind}(2021{\natexlab{a}})}]{susskind2021sitter}%
  \BibitemOpen
  \bibfield  {author} {\bibinfo {author} {\bibfnamefont {L.}~\bibnamefont
  {Susskind}},\ }\href@noop {} {\bibfield  {journal} {\bibinfo  {journal}
  {arXiv preprint arXiv:2106.03964}\ } (\bibinfo {year}
  {2021}{\natexlab{a}})}\BibitemShut {NoStop}%
\bibitem [{\citenamefont {Susskind}(2021{\natexlab{b}})}]{Susskind:2021esx}%
  \BibitemOpen
  \bibfield  {author} {\bibinfo {author} {\bibfnamefont {L.}~\bibnamefont
  {Susskind}},\ }\href@noop {} {\enquote {\bibinfo {title} {{Entanglement and
  Chaos in De Sitter Holography: An SYK Example}},}\ } (\bibinfo {year}
  {2021}{\natexlab{b}}),\ \Eprint {https://arxiv.org/abs/2109.14104}
  {arXiv:2109.14104 [hep-th]} \BibitemShut {NoStop}%
\bibitem [{\citenamefont {Dubovsky}, \citenamefont {Senatore},\ and\
  \citenamefont {Villadoro}(2009)}]{Dubovsky:2008rf}%
  \BibitemOpen
  \bibfield  {author} {\bibinfo {author} {\bibfnamefont {S.}~\bibnamefont
  {Dubovsky}}, \bibinfo {author} {\bibfnamefont {L.}~\bibnamefont {Senatore}},\
  and\ \bibinfo {author} {\bibfnamefont {G.}~\bibnamefont {Villadoro}},\ }\href
  {https://doi.org/10.1088/1126-6708/2009/04/118} {\bibfield  {journal}
  {\bibinfo  {journal} {JHEP}\ }\textbf {\bibinfo {volume} {04}},\ \bibinfo
  {pages} {118} (\bibinfo {year} {2009})},\ \Eprint
  {https://arxiv.org/abs/0812.2246} {arXiv:0812.2246 [hep-th]} \BibitemShut
  {NoStop}%
\bibitem [{\citenamefont {Chakraborty}\ \emph {et~al.}(2023)\citenamefont
  {Chakraborty}, \citenamefont {Chakravarty}, \citenamefont {Godet},
  \citenamefont {Paul},\ and\ \citenamefont {Raju}}]{Chakraborty:2023los}%
  \BibitemOpen
  \bibfield  {author} {\bibinfo {author} {\bibfnamefont {T.}~\bibnamefont
  {Chakraborty}}, \bibinfo {author} {\bibfnamefont {J.}~\bibnamefont
  {Chakravarty}}, \bibinfo {author} {\bibfnamefont {V.}~\bibnamefont {Godet}},
  \bibinfo {author} {\bibfnamefont {P.}~\bibnamefont {Paul}},\ and\ \bibinfo
  {author} {\bibfnamefont {S.}~\bibnamefont {Raju}},\ }\href@noop {} {\
  (\bibinfo {year} {2023})},\ \Eprint {https://arxiv.org/abs/2303.16316}
  {arXiv:2303.16316 [hep-th]} \BibitemShut {NoStop}%
\bibitem [{\citenamefont {Arad}, \citenamefont {Kuwahara},\ and\ \citenamefont
  {Landau}(2016)}]{Arad:2014znf}%
  \BibitemOpen
  \bibfield  {author} {\bibinfo {author} {\bibfnamefont {I.}~\bibnamefont
  {Arad}}, \bibinfo {author} {\bibfnamefont {T.}~\bibnamefont {Kuwahara}},\
  and\ \bibinfo {author} {\bibfnamefont {Z.}~\bibnamefont {Landau}},\ }\href
  {https://doi.org/10.1088/1742-5468/2016/03/033301} {\bibfield  {journal}
  {\bibinfo  {journal} {J. Stat. Mech.}\ }\textbf {\bibinfo {volume} {1603}},\
  \bibinfo {pages} {033301} (\bibinfo {year} {2016})},\ \Eprint
  {https://arxiv.org/abs/1406.3898} {arXiv:1406.3898 [quant-ph]} \BibitemShut
  {NoStop}%
\bibitem [{\citenamefont {Brand\~ao}\ \emph {et~al.}(2019)\citenamefont
  {Brand\~ao}, \citenamefont {Crosson}, \citenamefont {\ifmmode
  \mbox{\c{S}}\else \c{S}\fi{}ahino\ifmmode~\breve{g}\else \u{g}\fi{}lu},\ and\
  \citenamefont {Bowen}}]{Crosson}%
  \BibitemOpen
  \bibfield  {author} {\bibinfo {author} {\bibfnamefont {F.~G. S.~L.}\
  \bibnamefont {Brand\~ao}}, \bibinfo {author} {\bibfnamefont {E.}~\bibnamefont
  {Crosson}}, \bibinfo {author} {\bibfnamefont {M.~B.}\ \bibnamefont {\ifmmode
  \mbox{\c{S}}\else \c{S}\fi{}ahino\ifmmode~\breve{g}\else \u{g}\fi{}lu}},\
  and\ \bibinfo {author} {\bibfnamefont {J.}~\bibnamefont {Bowen}},\ }\href
  {https://doi.org/10.1103/PhysRevLett.123.110502} {\bibfield  {journal}
  {\bibinfo  {journal} {Phys. Rev. Lett.}\ }\textbf {\bibinfo {volume} {123}},\
  \bibinfo {pages} {110502} (\bibinfo {year} {2019})}\BibitemShut {NoStop}%
\bibitem [{\citenamefont {Cotler}\ and\ \citenamefont
  {Strominger}(2022)}]{Cotler:2022weg}%
  \BibitemOpen
  \bibfield  {author} {\bibinfo {author} {\bibfnamefont {J.}~\bibnamefont
  {Cotler}}\ and\ \bibinfo {author} {\bibfnamefont {A.}~\bibnamefont
  {Strominger}},\ }\href@noop {} {\  (\bibinfo {year} {2022})},\ \Eprint
  {https://arxiv.org/abs/2201.11658} {arXiv:2201.11658 [hep-th]} \BibitemShut
  {NoStop}%
\bibitem [{\citenamefont {Cotler}\ and\ \citenamefont
  {Jensen}(2023)}]{Cotler:2023eza}%
  \BibitemOpen
  \bibfield  {author} {\bibinfo {author} {\bibfnamefont {J.}~\bibnamefont
  {Cotler}}\ and\ \bibinfo {author} {\bibfnamefont {K.}~\bibnamefont
  {Jensen}},\ }\href@noop {} {\enquote {\bibinfo {title} {{Isometric evolution
  in de Sitter quantum gravity}},}\ } (\bibinfo {year} {2023}),\ \Eprint
  {https://arxiv.org/abs/2302.06603} {arXiv:2302.06603 [hep-th]} \BibitemShut
  {NoStop}%
\bibitem [{\citenamefont {H\"ohn}(2014)}]{Hohn:2014uvt}%
  \BibitemOpen
  \bibfield  {author} {\bibinfo {author} {\bibfnamefont {P.~A.}\ \bibnamefont
  {H\"ohn}},\ }\href {https://doi.org/10.1063/1.4890558} {\bibfield  {journal}
  {\bibinfo  {journal} {J. Math. Phys.}\ }\textbf {\bibinfo {volume} {55}},\
  \bibinfo {pages} {083508} (\bibinfo {year} {2014})},\ \Eprint
  {https://arxiv.org/abs/1401.6062} {arXiv:1401.6062 [gr-qc]} \BibitemShut
  {NoStop}%
\bibitem [{\citenamefont {Vidal}(2008)}]{Vidal:2008zz}%
  \BibitemOpen
  \bibfield  {author} {\bibinfo {author} {\bibfnamefont {G.}~\bibnamefont
  {Vidal}},\ }\href {https://doi.org/10.1103/PhysRevLett.101.110501} {\bibfield
   {journal} {\bibinfo  {journal} {Phys. Rev. Lett.}\ }\textbf {\bibinfo
  {volume} {101}},\ \bibinfo {pages} {110501} (\bibinfo {year} {2008})},\
  \Eprint {https://arxiv.org/abs/quant-ph/0610099} {arXiv:quant-ph/0610099}
  \BibitemShut {NoStop}%
\bibitem [{\citenamefont {Pfeifer}, \citenamefont {Evenbly},\ and\
  \citenamefont {Vidal}(2009)}]{Pfeifer:2009criticalMERA}%
  \BibitemOpen
  \bibfield  {author} {\bibinfo {author} {\bibfnamefont {R.~N.~C.}\
  \bibnamefont {Pfeifer}}, \bibinfo {author} {\bibfnamefont {G.}~\bibnamefont
  {Evenbly}},\ and\ \bibinfo {author} {\bibfnamefont {G.}~\bibnamefont
  {Vidal}},\ }\href {https://doi.org/10.1103/PhysRevA.79.040301} {\bibfield
  {journal} {\bibinfo  {journal} {Phys. Rev. A}\ }\textbf {\bibinfo {volume}
  {79}},\ \bibinfo {pages} {040301(R)} (\bibinfo {year} {2009})},\ \Eprint
  {https://arxiv.org/abs/0810.0580} {arXiv:0810.0580 [cond-mat.str-el]}
  \BibitemShut {NoStop}%
\bibitem [{\citenamefont {Evenbly}\ and\ \citenamefont
  {Vidal}(2009)}]{Evenbly:2007hxg}%
  \BibitemOpen
  \bibfield  {author} {\bibinfo {author} {\bibfnamefont {G.}~\bibnamefont
  {Evenbly}}\ and\ \bibinfo {author} {\bibfnamefont {G.}~\bibnamefont
  {Vidal}},\ }\href {https://doi.org/10.1103/PhysRevB.79.144108} {\bibfield
  {journal} {\bibinfo  {journal} {Phys. Rev. B}\ }\textbf {\bibinfo {volume}
  {79}},\ \bibinfo {pages} {144108} (\bibinfo {year} {2009})},\ \Eprint
  {https://arxiv.org/abs/0707.1454} {arXiv:0707.1454 [cond-mat.str-el]}
  \BibitemShut {NoStop}%
\bibitem [{\citenamefont {{Vidal}}(2010)}]{Vidal.ERintro}%
  \BibitemOpen
  \bibfield  {author} {\bibinfo {author} {\bibfnamefont {G.}~\bibnamefont
  {{Vidal}}},\ }\href@noop {} {\bibfield  {journal} {\bibinfo  {journal}
  {Understanding quantum phase transitions}\ } (\bibinfo {year}
  {2010})}\BibitemShut {NoStop}%
\bibitem [{\citenamefont {Swingle}(2012)}]{Swingle:2009bg}%
  \BibitemOpen
  \bibfield  {author} {\bibinfo {author} {\bibfnamefont {B.}~\bibnamefont
  {Swingle}},\ }\href {https://doi.org/10.1103/PhysRevD.86.065007} {\bibfield
  {journal} {\bibinfo  {journal} {Phys. Rev. D}\ }\textbf {\bibinfo {volume}
  {86}},\ \bibinfo {pages} {065007} (\bibinfo {year} {2012})},\ \Eprint
  {https://arxiv.org/abs/0905.1317} {arXiv:0905.1317 [cond-mat.str-el]}
  \BibitemShut {NoStop}%
\bibitem [{\citenamefont {Bao}\ \emph {et~al.}(2015)\citenamefont {Bao},
  \citenamefont {Cao}, \citenamefont {Carroll}, \citenamefont {Chatwin-Davies},
  \citenamefont {Hunter-Jones}, \citenamefont {Pollack},\ and\ \citenamefont
  {Remmen}}]{Bao:2015uaa}%
  \BibitemOpen
  \bibfield  {author} {\bibinfo {author} {\bibfnamefont {N.}~\bibnamefont
  {Bao}}, \bibinfo {author} {\bibfnamefont {C.}~\bibnamefont {Cao}}, \bibinfo
  {author} {\bibfnamefont {S.~M.}\ \bibnamefont {Carroll}}, \bibinfo {author}
  {\bibfnamefont {A.}~\bibnamefont {Chatwin-Davies}}, \bibinfo {author}
  {\bibfnamefont {N.}~\bibnamefont {Hunter-Jones}}, \bibinfo {author}
  {\bibfnamefont {J.}~\bibnamefont {Pollack}},\ and\ \bibinfo {author}
  {\bibfnamefont {G.~N.}\ \bibnamefont {Remmen}},\ }\href
  {https://doi.org/10.1103/PhysRevD.91.125036} {\bibfield  {journal} {\bibinfo
  {journal} {Phys. Rev. D}\ }\textbf {\bibinfo {volume} {91}},\ \bibinfo
  {pages} {125036} (\bibinfo {year} {2015})},\ \Eprint
  {https://arxiv.org/abs/1504.06632} {arXiv:1504.06632 [hep-th]} \BibitemShut
  {NoStop}%
\bibitem [{\citenamefont {Czech}\ \emph {et~al.}(2016)\citenamefont {Czech},
  \citenamefont {Lamprou}, \citenamefont {McCandlish},\ and\ \citenamefont
  {Sully}}]{czech2016tensor}%
  \BibitemOpen
  \bibfield  {author} {\bibinfo {author} {\bibfnamefont {B.}~\bibnamefont
  {Czech}}, \bibinfo {author} {\bibfnamefont {L.}~\bibnamefont {Lamprou}},
  \bibinfo {author} {\bibfnamefont {S.}~\bibnamefont {McCandlish}},\ and\
  \bibinfo {author} {\bibfnamefont {J.}~\bibnamefont {Sully}},\ }\href@noop {}
  {\bibfield  {journal} {\bibinfo  {journal} {Journal of High Energy Physics}\
  }\textbf {\bibinfo {volume} {2016}},\ \bibinfo {pages} {1} (\bibinfo {year}
  {2016})}\BibitemShut {NoStop}%
\bibitem [{\citenamefont {Beny}(2013)}]{Beny:2011vh}%
  \BibitemOpen
  \bibfield  {author} {\bibinfo {author} {\bibfnamefont {C.}~\bibnamefont
  {Beny}},\ }\href {https://doi.org/10.1088/1367-2630/15/2/023020} {\bibfield
  {journal} {\bibinfo  {journal} {New J. Phys.}\ }\textbf {\bibinfo {volume}
  {15}},\ \bibinfo {pages} {023020} (\bibinfo {year} {2013})},\ \Eprint
  {https://arxiv.org/abs/1110.4872} {arXiv:1110.4872 [quant-ph]} \BibitemShut
  {NoStop}%
\bibitem [{\citenamefont {Sinai~Kunkolienkar}\ and\ \citenamefont
  {Banerjee}(2017)}]{SinaiKunkolienkar:2016lgg}%
  \BibitemOpen
  \bibfield  {author} {\bibinfo {author} {\bibfnamefont {R.}~\bibnamefont
  {Sinai~Kunkolienkar}}\ and\ \bibinfo {author} {\bibfnamefont
  {K.}~\bibnamefont {Banerjee}},\ }\href
  {https://doi.org/10.1142/S0218271817501437} {\bibfield  {journal} {\bibinfo
  {journal} {Int. J. Mod. Phys. D}\ }\textbf {\bibinfo {volume} {26}},\
  \bibinfo {pages} {1750143} (\bibinfo {year} {2017})},\ \Eprint
  {https://arxiv.org/abs/1611.08581} {arXiv:1611.08581 [hep-th]} \BibitemShut
  {NoStop}%
\bibitem [{\citenamefont {Niermann}\ and\ \citenamefont
  {Osborne}(2022)}]{osborne}%
  \BibitemOpen
  \bibfield  {author} {\bibinfo {author} {\bibfnamefont {L.}~\bibnamefont
  {Niermann}}\ and\ \bibinfo {author} {\bibfnamefont {T.~J.}\ \bibnamefont
  {Osborne}},\ }\href {https://doi.org/10.1103/PhysRevD.105.125009} {\bibfield
  {journal} {\bibinfo  {journal} {Phys. Rev. D}\ }\textbf {\bibinfo {volume}
  {105}},\ \bibinfo {pages} {125009} (\bibinfo {year} {2022})}\BibitemShut
  {NoStop}%
\bibitem [{\citenamefont {{Milsted}}\ and\ \citenamefont
  {{Vidal}}(2018)}]{Milsted}%
  \BibitemOpen
  \bibfield  {author} {\bibinfo {author} {\bibfnamefont {A.}~\bibnamefont
  {{Milsted}}}\ and\ \bibinfo {author} {\bibfnamefont {G.}~\bibnamefont
  {{Vidal}}},\ }\href {https://doi.org/10.48550/arXiv.1812.00529} {\bibfield
  {journal} {\bibinfo  {journal} {arXiv e-prints}\ ,\ \bibinfo {eid}
  {arXiv:1812.00529}} (\bibinfo {year} {2018})},\ \Eprint
  {https://arxiv.org/abs/1812.00529} {arXiv:1812.00529 [hep-th]} \BibitemShut
  {NoStop}%
\bibitem [{\citenamefont {Witten}(2001)}]{Witten:2001kn}%
  \BibitemOpen
  \bibfield  {author} {\bibinfo {author} {\bibfnamefont {E.}~\bibnamefont
  {Witten}},\ }in\ \href@noop {} {\emph {\bibinfo {booktitle} {{Strings 2001:
  International Conference}}}}\ (\bibinfo {year} {2001})\ \Eprint
  {https://arxiv.org/abs/hep-th/0106109} {arXiv:hep-th/0106109} \BibitemShut
  {NoStop}%
\bibitem [{\citenamefont {Bousso}(2000)}]{Bousso:2000nf}%
  \BibitemOpen
  \bibfield  {author} {\bibinfo {author} {\bibfnamefont {R.}~\bibnamefont
  {Bousso}},\ }\href {https://doi.org/10.1088/1126-6708/2000/11/038} {\bibfield
   {journal} {\bibinfo  {journal} {JHEP}\ }\textbf {\bibinfo {volume} {11}},\
  \bibinfo {pages} {038} (\bibinfo {year} {2000})},\ \Eprint
  {https://arxiv.org/abs/hep-th/0010252} {arXiv:hep-th/0010252} \BibitemShut
  {NoStop}%
\bibitem [{\citenamefont {Parikh}\ and\ \citenamefont
  {Verlinde}(2005)}]{Parikh:2004wh}%
  \BibitemOpen
  \bibfield  {author} {\bibinfo {author} {\bibfnamefont {M.~K.}\ \bibnamefont
  {Parikh}}\ and\ \bibinfo {author} {\bibfnamefont {E.~P.}\ \bibnamefont
  {Verlinde}},\ }\href {https://doi.org/10.1088/1126-6708/2005/01/054}
  {\bibfield  {journal} {\bibinfo  {journal} {JHEP}\ }\textbf {\bibinfo
  {volume} {01}},\ \bibinfo {pages} {054} (\bibinfo {year} {2005})},\ \Eprint
  {https://arxiv.org/abs/hep-th/0410227} {arXiv:hep-th/0410227} \BibitemShut
  {NoStop}%
\bibitem [{\citenamefont {Dyson}, \citenamefont {Kleban},\ and\ \citenamefont
  {Susskind}(2002)}]{Dyson:2002pf}%
  \BibitemOpen
  \bibfield  {author} {\bibinfo {author} {\bibfnamefont {L.}~\bibnamefont
  {Dyson}}, \bibinfo {author} {\bibfnamefont {M.}~\bibnamefont {Kleban}},\ and\
  \bibinfo {author} {\bibfnamefont {L.}~\bibnamefont {Susskind}},\ }\href
  {https://doi.org/10.1088/1126-6708/2002/10/011} {\bibfield  {journal}
  {\bibinfo  {journal} {JHEP}\ }\textbf {\bibinfo {volume} {10}},\ \bibinfo
  {pages} {011} (\bibinfo {year} {2002})},\ \Eprint
  {https://arxiv.org/abs/hep-th/0208013} {arXiv:hep-th/0208013} \BibitemShut
  {NoStop}%
\bibitem [{\citenamefont {Banks}(2001)}]{Banks:2000fe}%
  \BibitemOpen
  \bibfield  {author} {\bibinfo {author} {\bibfnamefont {T.}~\bibnamefont
  {Banks}},\ }\href {https://doi.org/10.1142/S0217751X01003998} {\bibfield
  {journal} {\bibinfo  {journal} {Int. J. Mod. Phys. A}\ }\textbf {\bibinfo
  {volume} {16}},\ \bibinfo {pages} {910} (\bibinfo {year} {2001})},\ \Eprint
  {https://arxiv.org/abs/hep-th/0007146} {arXiv:hep-th/0007146} \BibitemShut
  {NoStop}%
\bibitem [{\citenamefont {Shaghoulian}(2022)}]{Shaghoulian:2021cef}%
  \BibitemOpen
  \bibfield  {author} {\bibinfo {author} {\bibfnamefont {E.}~\bibnamefont
  {Shaghoulian}},\ }\href {https://doi.org/10.1007/JHEP01(2022)132} {\bibfield
  {journal} {\bibinfo  {journal} {JHEP}\ }\textbf {\bibinfo {volume} {01}},\
  \bibinfo {pages} {132} (\bibinfo {year} {2022})},\ \Eprint
  {https://arxiv.org/abs/2110.13210} {arXiv:2110.13210 [hep-th]} \BibitemShut
  {NoStop}%
\bibitem [{\citenamefont {Chandrasekaran}\ \emph {et~al.}(2022)\citenamefont
  {Chandrasekaran}, \citenamefont {Longo}, \citenamefont {Penington},\ and\
  \citenamefont {Witten}}]{chandrasekaran2022algebra}%
  \BibitemOpen
  \bibfield  {author} {\bibinfo {author} {\bibfnamefont {V.}~\bibnamefont
  {Chandrasekaran}}, \bibinfo {author} {\bibfnamefont {R.}~\bibnamefont
  {Longo}}, \bibinfo {author} {\bibfnamefont {G.}~\bibnamefont {Penington}},\
  and\ \bibinfo {author} {\bibfnamefont {E.}~\bibnamefont {Witten}},\
  }\href@noop {} {\bibfield  {journal} {\bibinfo  {journal} {arXiv preprint
  arXiv:2206.10780}\ } (\bibinfo {year} {2022})}\BibitemShut {NoStop}%
\bibitem [{\citenamefont {Ji}, \citenamefont {Liu},\ and\ \citenamefont
  {Song}(2018)}]{JiLiuSong}%
  \BibitemOpen
  \bibfield  {author} {\bibinfo {author} {\bibfnamefont {Z.}~\bibnamefont
  {Ji}}, \bibinfo {author} {\bibfnamefont {Y.-K.}\ \bibnamefont {Liu}},\ and\
  \bibinfo {author} {\bibfnamefont {F.}~\bibnamefont {Song}},\ }in\ \href@noop
  {} {\emph {\bibinfo {booktitle} {Advances in Cryptology -- CRYPTO 2018}}},\
  \bibinfo {editor} {edited by\ \bibinfo {editor} {\bibfnamefont
  {H.}~\bibnamefont {Shacham}}\ and\ \bibinfo {editor} {\bibfnamefont
  {A.}~\bibnamefont {Boldyreva}}}\ (\bibinfo  {publisher} {Springer
  International Publishing},\ \bibinfo {address} {Cham},\ \bibinfo {year}
  {2018})\ pp.\ \bibinfo {pages} {126--152}\BibitemShut {NoStop}%
\bibitem [{\citenamefont {Bouland}\ \emph {et~al.}(2022)\citenamefont
  {Bouland}, \citenamefont {Fefferman}, \citenamefont {Ghosh}, \citenamefont
  {Vazirani},\ and\ \citenamefont {Zhou}}]{Bouland:2022ovo}%
  \BibitemOpen
  \bibfield  {author} {\bibinfo {author} {\bibfnamefont {A.}~\bibnamefont
  {Bouland}}, \bibinfo {author} {\bibfnamefont {B.}~\bibnamefont {Fefferman}},
  \bibinfo {author} {\bibfnamefont {S.}~\bibnamefont {Ghosh}}, \bibinfo
  {author} {\bibfnamefont {U.}~\bibnamefont {Vazirani}},\ and\ \bibinfo
  {author} {\bibfnamefont {Z.}~\bibnamefont {Zhou}},\ }\href@noop {} {\
  (\bibinfo {year} {2022})},\ \Eprint {https://arxiv.org/abs/2211.00747}
  {arXiv:2211.00747 [quant-ph]} \BibitemShut {NoStop}%
\bibitem [{\citenamefont {Evenbly}\ and\ \citenamefont
  {White}(2016)}]{Evenbly:2016cly}%
  \BibitemOpen
  \bibfield  {author} {\bibinfo {author} {\bibfnamefont {G.}~\bibnamefont
  {Evenbly}}\ and\ \bibinfo {author} {\bibfnamefont {S.~R.}\ \bibnamefont
  {White}},\ }\href {https://doi.org/10.1103/PhysRevLett.116.140403} {\bibfield
   {journal} {\bibinfo  {journal} {Phys. Rev. Lett.}\ }\textbf {\bibinfo
  {volume} {116}},\ \bibinfo {pages} {140403} (\bibinfo {year} {2016})},\
  \Eprint {https://arxiv.org/abs/1602.01166} {arXiv:1602.01166
  [cond-mat.str-el]} \BibitemShut {NoStop}%
\bibitem [{\citenamefont {Jahn}\ \emph {et~al.}(2019)\citenamefont {Jahn},
  \citenamefont {Gluza}, \citenamefont {Pastawski},\ and\ \citenamefont
  {Eisert}}]{Jahn:2017tls}%
  \BibitemOpen
  \bibfield  {author} {\bibinfo {author} {\bibfnamefont {A.}~\bibnamefont
  {Jahn}}, \bibinfo {author} {\bibfnamefont {M.}~\bibnamefont {Gluza}},
  \bibinfo {author} {\bibfnamefont {F.}~\bibnamefont {Pastawski}},\ and\
  \bibinfo {author} {\bibfnamefont {J.}~\bibnamefont {Eisert}},\ }\href
  {https://doi.org/10.1126/sciadv.aaw0092} {\bibfield  {journal} {\bibinfo
  {journal} {Sci. Adv.}\ }\textbf {\bibinfo {volume} {5}},\ \bibinfo {pages}
  {eaaw0092} (\bibinfo {year} {2019})},\ \Eprint
  {https://arxiv.org/abs/1711.03109} {arXiv:1711.03109 [quant-ph]} \BibitemShut
  {NoStop}%
\bibitem [{\citenamefont {Ryu}\ and\ \citenamefont
  {Takayanagi}(2006)}]{Ryu:2006bv}%
  \BibitemOpen
  \bibfield  {author} {\bibinfo {author} {\bibfnamefont {S.}~\bibnamefont
  {Ryu}}\ and\ \bibinfo {author} {\bibfnamefont {T.}~\bibnamefont
  {Takayanagi}},\ }\href {https://doi.org/10.1103/PhysRevLett.96.181602}
  {\bibfield  {journal} {\bibinfo  {journal} {Phys. Rev. Lett.}\ }\textbf
  {\bibinfo {volume} {96}},\ \bibinfo {pages} {181602} (\bibinfo {year}
  {2006})},\ \Eprint {https://arxiv.org/abs/hep-th/0603001}
  {arXiv:hep-th/0603001} \BibitemShut {NoStop}%
\bibitem [{\citenamefont {Hayden}\ \emph {et~al.}(2016)\citenamefont {Hayden},
  \citenamefont {Nezami}, \citenamefont {Qi}, \citenamefont {Thomas},
  \citenamefont {Walter},\ and\ \citenamefont {Yang}}]{Hayden:2016cfa}%
  \BibitemOpen
  \bibfield  {author} {\bibinfo {author} {\bibfnamefont {P.}~\bibnamefont
  {Hayden}}, \bibinfo {author} {\bibfnamefont {S.}~\bibnamefont {Nezami}},
  \bibinfo {author} {\bibfnamefont {X.-L.}\ \bibnamefont {Qi}}, \bibinfo
  {author} {\bibfnamefont {N.}~\bibnamefont {Thomas}}, \bibinfo {author}
  {\bibfnamefont {M.}~\bibnamefont {Walter}},\ and\ \bibinfo {author}
  {\bibfnamefont {Z.}~\bibnamefont {Yang}},\ }\href
  {https://doi.org/10.1007/JHEP11(2016)009} {\bibfield  {journal} {\bibinfo
  {journal} {JHEP}\ }\textbf {\bibinfo {volume} {11}},\ \bibinfo {pages} {009}
  (\bibinfo {year} {2016})},\ \Eprint {https://arxiv.org/abs/1601.01694}
  {arXiv:1601.01694 [hep-th]} \BibitemShut {NoStop}%
\end{thebibliography}
\end{document}